\documentclass[british,english,manuscript]{aastex6}
\usepackage[T1]{fontenc}
\usepackage[latin9]{inputenc}
\usepackage{refstyle}
\usepackage{float}
\usepackage{booktabs}
\usepackage{amsbsy}
\usepackage{amstext}
\usepackage{amssymb}
\usepackage{graphicx}

\makeatletter

\AtBeginDocument{\providecommand\subsecref[1]{\ref{subsec:#1}}}
\AtBeginDocument{\providecommand\secref[1]{\ref{sec:#1}}}
\AtBeginDocument{\providecommand\eqref[1]{\ref{eq:#1}}}
\AtBeginDocument{\providecommand\figref[1]{\ref{fig:#1}}}
\AtBeginDocument{\providecommand\tabref[1]{\ref{tab:#1}}}
\providecommand{\tabularnewline}{\\}
\newcommand{\lyxdot}{.}

\RS@ifundefined{subsecref}
  {\newref{subsec}{name = \RSsectxt}}
  {}
\RS@ifundefined{thmref}
  {\def\RSthmtxt{theorem~}\newref{thm}{name = \RSthmtxt}}
  {}
\RS@ifundefined{lemref}
  {\def\RSlemtxt{lemma~}\newref{lem}{name = \RSlemtxt}}
  {}

\usepackage{paralist}

\slugcomment{}
\shorttitle{}
\shortauthors{}

\renewcommand{\tabref}{\Tabref}
\renewcommand{\figref}{\Figref}

\usepackage{amsmath}
\usepackage{cleveref}

\makeatother

\usepackage{babel}
\begin{document}

\title{Pitch-Angle Diffusion and Bohm-type Approximations\\
 in Diffusive Shock Acceleration}

\author{John Daniel Riordan\altaffilmark{1}, Asaf Pe'er\altaffilmark{1,2}}

\altaffiltext{1}{Department of Physics, University College Cork, Cork, Ireland}

\altaffiltext{2}{Department of Physics, Bar-Ilan University, Ramat-Gan, 52900, Israel}

\email{dannyriordan@gmail.com}
\begin{abstract}
The problem of accelerating cosmic rays is one of fundamental importance,
particularly given the uncertainty in the conditions inside the acceleration
sites. Here we examine Diffusive Shock Acceleration in arbitrary turbulent
magnetic fields, constructing a new model that is capable of bridging
the gap between the very weak ($\delta B/B_{0}\ll1$) and the strong
turbulence regimes. To describe the diffusion we provide quantitative
analytical description of the \textquotedbl{}Bohm exponent\textquotedbl{}
in each regime. We show that our results converge to the well known
quasi-linear theory in the weak turbulence regime. In the strong regime,
we quantify the limitations of the Bohm-type models. Furthermore,
our results account for the anomalous diffusive behaviour which has
been noted previously. Finally, we discuss the implications of our
model in the study of possible acceleration sites in different astronomical
objects.
\end{abstract}


\section{Introduction}

Diffusive Shock Acceleration (DSA), also known as First-Order Fermi
Acceleration is a leading model in explaining the acceleration of
particles and production of cosmic rays (CRs) in various astronomical
objects \citep{Fermi1949,Bell1978,Blandford1987,Ellison1990,Malkov1997}.
In this model particles may gain energy by repeatedly crossing a shock
wave by elastically reflecting from magnetic turbulence on each side,
producing an $E^{-2}$ energy spectrum sonsistent with CR particle
observations on earth (for details see \citet{Blandford1987,Dermer2009}
and references therein). 

This model has been extensively studied in the past (see \citet{Blandford1987}
for a review) and is well understood in the regime in which the following
two conditions hold:
\begin{inparaenum}[1)]
\item weak turbulence, namely $\delta B/B_{0}\ll1$, and 
\item the test-particle approximation. 
\end{inparaenum}
Here $B_{0}$ is the magnitude of a guiding magnetic field that exists
in the shock vicinity, and $\delta B$ is the magnitude of a turbulent
field. The ``test-particle approximation'' implies that the fraction
of energy carried by the accelerated particles is negligible with
respect to the thermal energy of the plasma, hence these particles
do not contribute significantly to the turbulence. In weak turbulence
this is a natural assumption, since the magnetic energy available
for acceleration is small. Conversely if the test particle approximation
is valid the turbulence generated by the accelerated particles is
negligible. For this reason the test-particle approximation is typically
employed unless strong turbulence is present.

While it is widely believed that these conditions are met in sources
that are likely responsible for acceleration of CRs up to the observed
``knee'' in the CR spectrum ($\approx10^{15}\mathrm{eV}$) \citep{Lagage83,Volk1988,Bell2014},
it is far from being clear whether these conditions are met in sources
that accelerate CRs to higher energies \citep{Lucek2000,Achterberg2001}.
Furthermore, as has been shown by \citet{Bykov2014}, in order to
generate sufficient turbulent magnetic fields necessary for reflecting
the particles back and forth across the shock, the self-generated
turbulence of the accelerated particles must be treated. As we will
explain, current diffusion models make assumptions that may not be
valid as $\delta B/B_{0}$ increases due to these self-generated waves.

Previous studies of DSA can be broadly divided into three categories.
The first is the Semi-Analytic approach (e.g. \citet{Kirk1989,Malkov1997,Amato2005,Caprioli2010a}),
in which the particles are described in terms of distribution functions,
enabling analytic or numerical solution of the transport equations.
While this is the fastest method, reliable models only exist in a
very limited parameter range (weak turbulence, small-angle scattering,
weakly anisotropic, etc.). Furthermore a heuristic prescription for
the diffusion is required. The second is the\textbf{ }Monte-Carlo
approach (e.g. \citet{Ellison1990,Achterberg2001,Ellison2002,Summerlin2011,Bykov2017}),
in which the trajectories and properties of representative particles
are tracked and the average background magnetic fields are estimated.
The advantage of this method is that it enables the study of a large
parameter space region, and is very fast and therefore can be used
to track the particle trajectories over the entire region where the
acceleration is believed to occur \citep{Ellison2013}. On the other
hand this method uses simplifying assumptions about the structure
of the magnetic fields and the details of their interaction with the
particles. For example, several existing Monte-Carlo codes \citep{Achterberg2001,Vladimirov06}
use scattering models which are either limited to weak turbulence,
such as quasilinear theory (QLT; see \citet{Jokipii1966,Shalchi2009c}
and further discussion below), or are not well supported theoretically,
such as the Bohm type \citep{Casse2002}. The third approach is Particle-In-Cell
(PIC) simulations \citep{Birdsall1985,Silva03,Frederiksen2004a,Spitkovsky2008,Sironi2011,Guo2014,Bai2014}.
These codes simultaneously solve for particle trajectories and electromagnetic
fields in a fully self-consistent way. They therefore provide full
treament of particle acceleration, magnetic turbulence and formation
of shocks. However, existing codes are prohibitevely expensive computationally
and are therefore limited to very small ranges in time and space,
typically many orders of magnitude less than the regime in which particles
are believed to be accelerated \citep{Vladimirov2009a}.

Of the three approaches the one that currently seems best applicable
to astrophysical environments is the Monte-Carlo approach. Analytic
techniques quickly become unwieldy when trying to account for e.g.
strong turbulence, oblique shocks or plasma instabilites which develop
under different conditions \citep{Caprioli2010,Summerlin2011}. On
the other hand, the computational power required for carrying out
a PIC simulation over the full dynamical range is not expected to
be available for many years. While the Monte-Carlo approach also suffers
several weaknesses as described above, some of these weakness can
be treated with reasonable computational time. 

At the heart of the Monte-Carlo approach lies a description of the
particle-field interaction. As described above various authors use
various prescriptions (e.g. \citet{Vladimirov06,Vladimirov2008,Tautz2013}),
which rely on very different assumptions. The purpose of the current
work is to examine and quantify the validity of the two most frequently
used of these assumptions in describing the particle-field interactions
in Monte-Carlo codes, namely QLT and Bohm diffusion. As we will show
below, the results of the QLT approximation are sensitive to the timescale
over which the diffusion is measured. Furthermore Bohm diffusion does
not apply before the turbulence is very strong. Our results are therefore
relevant to the production of more accurate Monte-Carlo models in
the future.

Monte-Carlo codes (e.g. \citet{Ellison1990,Achterberg2001,Ellison2002,Summerlin2011,Bykov2017})
typically consider an idealised scenario, where energy changes and
local spatial variations are neglected. In such an environment the
wave-particle interaction is determined by a single quantity, the
particle's\emph{ pitch angle} $\vartheta$, i.e. the angle its velocity
vector makes with the direction of the background field. It is useful
to examine the stochastic behaviour of $\mu=\cos\vartheta$ as the
particle undergoes ``scattering'' from the magnetic turbulence.
Studies of this type, \emph{pitch-angle scattering} (e.g. \citet{Qin2009}),
typically treat this pitch angle as undergoing a random walk, being
``scattered'' each time its direction is rotated by interacting
with a turbulent wave. Analytic work has mainly centred on the ``quasilinear''
family of approximations, originally formulated by \citet{Jokipii1966},
in which the deviation from helical orbits is treated perturbatively
(see e.g. \citet{Schlickeiser2002,Shalchi2009c}) by averaging out
wave contributions over many gyrotimes. Furthermore it is useful to
consider the separate components of diffusion in the directions parallel
and perpendicular to the shock, since the latter is directly responsible
for the particle repeatedly crossing the shock and gaining energy
\citep{Shalchi2009c}. How this relates to pitch-angle scattering
is governed by the obliquity angle made between the background magnetic
field with the shock normal. Since in strong turbulence the influence
of the background field is less significant, we simplify our results
by averaging over pitch-angle, however we note that a separate treatment
of parallel and perpendicular diffusion may provide a more complete
description of the system \citep{Ferrand2014,Shalchi2015}.

The strength of the turbulent contribution is quantified by the \emph{turbulence
ratio} $\delta B/B_{0}$. We distinguish weak, intermediate and strong
turbulence as $\delta B/B_{0}\approx0$, $\lesssim1$ and $>1$ respectively.
The classical quasilinear approach requires a first-order approximation
in $\delta B/B_{0}$ around $0$. It has been shown to give an accurate
description of particle motion in the weak regime and various modifications
exist to extend its range to intermediate turbulence by including
higher-order terms \citep{Blandford1987,Schlickeiser2002,Shalchi2009c}.
It has been shown, however, in heliospheric observation \citep{Tu1995},
and at Saturn's bow shock \citep{Masters} that $\delta B/B_{0}$
can be as high as order unity. This turbulence level is also seen
in the numerical results of PIC simulations \citep{Sironi2011}. 

Additionally, QLT approximations result in a resonance condition,
in which the particles interact only with a resonant portion of the
magnetic turbulent wave spectrum ($k\approx1/r_{g}$ for wavenumber
$k$ and gyroradius $r_{g}$), though this is not necessarily the
case \citep{Li1997}. Furthermore in its original form QLT exhibits
the ``$90\textdegree$ problem'', in which particles with $\mu=0$
experience no scattering, in conflict with Monte-Carlo simulations
which do not use a scattering approximation \citep{Shalchi2005,Qin2009}.
This is due to second-order approximations when calculating velocity
and magnetic field correlations \citep{Jokipii1972,Giacalone1999,Tautz2010a}.
There are several extensions to QLT which address this problem by
adding nonlinear terms \citet{Shalchi2009c}, notably Second-Order
QLT (SOQLT). While it is valid in a larger turbulence range than QLT,
and remedies the ``$90\textdegree$ problem'', SOQLT still cannot
extend to intermediate turbulence and relies on a similar resonance
approximation.

Many large scale Monte-Carlo simulations, such as those presented
in e.g. \citet{Caprioli2010,Ellison2013}, out of computational necessity
instead treat interaction with the turbulence using a different pitch-angle
scattering model, namely the \emph{Bohm Diffusion} approximation.
In this approximation the particle's motion is described as undergoing
a series of discrete, isotropic  scatterings. In contrast to QLT
this approach does not represent resonant interaction with individual
waves or account for pitch-angle dependence of scatterings. Rather,
in the Bohm model, the mean free path $\lambda_{\mathrm{mfp}}$ between
scattering assumes the form 
\begin{equation}
\lambda_{\mathrm{mfp}}=\eta r_{g}^{\alpha}\label{eq:bohm-like form}
\end{equation}
 where the Bohm exponent $\alpha$ is a free parameter whose value
is unknown and is often taken as unity \citep{Baring2009} and $\eta$
is a coupling constant (see discussion in \subsecref{Validity-of-Bohm}).
This model was initially formulated in the context of electric field
interactions in laboratory plasmas \citep{Bohm1949} and is frequently
employed as a heuristic in astrophysics. There is some numerical
support for its validity in the context of DSA. \citet{Casse2002}
found it only to be valid when $\delta B\approx B_{0}$ (intermediate
turbulence) and $0.1<r_{g}k_{\mathrm{min}}<1$ where $k_{\mathrm{min}}$
is the smallest wavenumber in the turbulence for a power law spectrum,
despite the fact that Bohm diffusion is typically not assumed to rely
on resonant effects. \citet{Reville2008} find Bohm diffusion as an
upper limit when propagating particles of different energies against
a magnetic background obtained from MHD simulations. Other works
on pitch angle scattering have examined $\lambda_{\mathrm{mfp}}$
as a function of $\delta B/B_{0}$ , and concluded that this form
is valid in the strong turbulence region $\delta B\gtrsim B_{0}$
\citep{Shalchi2009b,Hussein2014}. 

In this work we aim to examine and quantify the limitations of the
QLT and Bohm approximations with the goal of better understanding
the wave-particle interactions involved in DSA. As we show below,
while QLT provides a good approximation up to intermediate turbulence,
this result is sensitive to the calculation method of diffusion coefficient,
and the timescale over which diffusion is measured. As for Bohm diffusion,
the classical $\alpha=1$ approximation is the correct asymptotic
solution at high turbulence but is of very limited validity. We therefore
provide the turbulence-dependent value of $\alpha$ as a function
$\delta B/B_{0}$ for a set of representative parameters, which smoothly
interpolates between the weak and strong limits. This extended Bohm-type
model will improve future Monte-Carlo simulations by accurately and
consistently modelling scattering across all levels of turbulence.
This paper is organised as follows. In \secref{Model} we describe
our model setup and computational methods. In \secref{Diffusion}
we discuss the pitch-angle diffusion coefficient $D_{\mu\mu}$. Our
results are presented in \secref{Results}. We discuss our findings
in \secref{Discussion}, before summarising and concluding in \secref{Conclusions}.

\section{Model and Methods}

\label{sec:Model}

We model the acceleration region as a three-dimensional collisionless
plasma adjacent to a shock front. We make no assumption with regard
to the direction of propagation. This plasma consists of a population
of identical charged particles, and a magentic field consisting of
both a uniform guiding component $\boldsymbol{B}_{0}\parallel\boldsymbol{z}$
and a turbulent component described by a population of Alfvén waves.
We further assume spatial homogeneity and cylindrical symmetry around
the $z$ direction. Since we expect particle acceleration to occur
in the vicinity of collisionless shocks we neglect electric fields
and Coulomb collisions \citep{Bret2018}. The particle's pitch angle
cosine is then given by $\mu=\frac{v_{z}}{v}$ where $\boldsymbol{v}$
is the particle's velocity. The particles propagate subject to the
Lorentz force and we track their trajectories and measure their collective
properties.

\subsection{Simulation\label{subsec:Simulation}}

For simulating the plasma system evolution a new simulation code has
been developed. This code is distinct in its ``wave population''
treatment of the magnetic field. The magnetic fields  are calculated
at every timestep at the particle's current spatial location, rather
than being evaluated on a grid at the beginning of the simulation
(as in previous studies of this type e.g. \citet{Giacalone1999,Mace2000,Reville2008,Tautz2010}).
 The advantages of this method are that 1) it makes the spatial resolution
effectively continuous and 2) it facilitates modifying the turbulence
spectrum during particle motion. The disadvantage is the higher cost
of performance.  Initial populations of waves (see \subsecref{Waves})
and particles are prescribed and the total magnetic field $\boldsymbol{B}$
is calculated as a function of position by summing the contribution
$\delta\boldsymbol{B}$ of each wave, along with the background field
$\boldsymbol{B}_{0}$. The system is evolved using the Newton-Lorentz
equation for particles of mass $m$ and charge $e$,
\begin{equation}
\frac{\mathrm{d}p^{\mu}}{\mathrm{d}\tau}=eF^{\mu\nu}u_{\nu}\label{eq:lorentz-cov}
\end{equation}
where $p^{\mu}=mu^{\mu}$ is the four momentum, $u^{\mu}=\gamma\left(c,\boldsymbol{v}\right)$
is the four-velocity, $\gamma$ is the Lorentz factor, $F^{\mu\nu}$
is the Maxwell tensor, and $\tau=t/\gamma$ is the proper time. Since
the force acting on the particle is assumed to be purely magnetic,
its energy $p^{0}=\gamma mc$ is conserved and the spatial part of
the above equation becomes:
\begin{equation}
\frac{\mathrm{d}^{2}\boldsymbol{x}}{\mathrm{d}t^{2}}=\frac{e}{m\gamma}\left(\boldsymbol{v}\boldsymbol{\times}\boldsymbol{B}\right)\label{eq:lorentznocov}
\end{equation}
 The particle trajectories are solved for using \eqref{lorentznocov}
and recorded in order to measure the diffusion coefficients (see \secref{Diffusion}).

The gyrotime $t_{g}$, angular gyrofrequency $\omega_{g}$, and gyroradius
$r_{g}$ respectively are defined as follows:

\begin{align}
t_{g} & =\frac{2\pi\gamma m}{eB_{\perp}}=\text{\ensuremath{\frac{2\pi}{\omega_{g}}}}\text{,}\\
r_{g} & =v_{\perp}/\omega_{g}=\left(1-\mu^{2}\right)^{1/2}v\frac{t_{g}}{2\pi}\text{,}
\end{align}
with $B_{\perp}$ the component of $\boldsymbol{B}$ perpendicular
to $\boldsymbol{v}$ and vice versa. When $t_{g}$ and $r_{g}$ are
used to normalise other quantities we take their values assuming $\boldsymbol{B}=\boldsymbol{B_{0}}$,
$\gamma=1$ and $\mu=0$, although these are both generally dependent
variables. In this work only non-relativistic values for $v$ are
taken. This is done for computational simplicity but should not qualitatively
affect the results (see discussion in \secref{Discussion}).

In running the simulation we normalise the speed of light, elementary
charge, proton mass and guiding magnetic field strength $\boldsymbol{B}_{0}$.
For this reason results are given in terms of $t_{g}$ and are applicable
to any magnetic field with suitable scaling of the time.

\subsection{Modelling the Turbulent Magnetic Field\label{subsec:Waves}}

The overall magnetic field comprises a constant background field $\boldsymbol{B}_{0}$
and turbulent field $\boldsymbol{\delta B}$. The turbulent field
is found by summing over a discrete population of waves at each position
$x^{j}$, as follows,

\begin{equation}
\boldsymbol{\delta B}=\sum_{\mathrm{waves}}A_{k}e^{i\left(k_{j}x^{j}+\phi_{k}\right)}\mathbf{n}\text{.}\label{eq:dBaswaves}
\end{equation}
Here $A_{k}$ is the amplitude of the wave with wavenumber $k$, $k_{i}$
is its wavevector, $\mathbf{n}$ is its polarisation vector and $\phi_{k}$
is its phase; latin indices run over spatial coordinates. The phases
and polarisation angle are chosen randomly from a uniform distribution
on $[0,2\pi]$\footnote{We note that the phase distribution may be non-uniform in strong turbulence
\citep{Blandford1987}}.

We distinguish waves having $\boldsymbol{k}=\left(0,0,k_{\parallel}\right)$
as \emph{slab} waves and $\boldsymbol{k}=k_{\perp}\left(\cos\vartheta_{\perp},\sin\vartheta_{\perp},0\right)$
for some angle $\vartheta_{\perp}$, as \emph{$2d$}-waves. Turbulence
containing both kinds of wave is said to be \emph{composite}. The
$2d$ waves can be further divided into \emph{full-$2d$} \citep{Shalchi2008a}
with  $\boldsymbol{\delta B}\perp\boldsymbol{B}_{0}$ or the more
general \emph{omnidirectional} type with another angle $\psi$ so
that $\boldsymbol{\delta B}\boldsymbol{\cdot}\boldsymbol{B}_{0}\propto\sin\psi$.
Observation of the solar environment suggests that full-$2d$ waves
may be a suitable model \citep{Bieber1996}. It has been proposed
these full-\emph{$2d$} waves represent ``magnetostatic structures''
\citep{Gray1996}. On the other hand, numerical simulations (e.g.
\citet{Bell2004,Reville2008}) have shown that waves with $\boldsymbol{\delta B}\boldsymbol{\cdot}\boldsymbol{B}_{0}\neq0$
may result from plasma instabilities at acceleration sites and therefore
omnidirectional waves must be used. The proportion of full-$2d$ turbulence
decreases with turbulence level, such that highly turbulent plasmas
tend towards isotropy \citep{Bell2004}, since Alfvén waves propagate
along the direction of the local $B$-field (which is $\approx\boldsymbol{B}_{0}$
only in the low turbulence case). For simplicity, in this work we
restrict our attention to composite turbulence comprising slab waves
and full-$2d$ waves, and defer omnidirectional turbulence to a future
work. 

The ratio of energies in each wave type, is parametrised by the \emph{slab
fraction} \citep{Bieber1996},

\begin{equation}
r_{\mathrm{slab}}=\frac{\delta B_{\text{slab }}^{2}}{\delta B_{\text{slab }}^{2}+\delta B_{\text{2d }}^{2}}=\frac{\delta B_{\text{slab }}^{2}}{\delta B_{\text{total }}^{2}}
\end{equation}
 for which a range of values is possible \citep{Tautz2011}. Here
$\delta B_{\text{slab }}^{2}$ and $\delta B_{\text{2d }}^{2}$ denote
the total energy in the slab and $2d$ waves respectively. Solar wind
observations give a value of $\approx0.2$ \citep{Gray1996,Shalchi2008a}
and in the presence of Bell instability the slab fraction saturates
at $\approx0.5$ \citep{Bell2004,Reville2008}.

In the present work for the purposes of simplicity it is assumed the
waves are static in time. This approximation is valid as long as $v_{A}\ll v$,
where wave propagate at the (nonrelativistic) Alfvén velocity $v_{A}=\frac{B_{0}}{\sqrt{\mu_{0}\rho}}$
, $\rho$ is the density of charge carriers, $\mu_{0}$ is the magnetic
permeability of the vacuum, and $v$ is particle velocity. This removes
the time dependence of the turbulence due to wave propagation and
the associated electric field (the electric component of an Alfvén
wave has magnitude $\sim v_{A}/c$). This assumption is valid for
plasmas that are not highly magnetised, but may be violated once the
Alfvén speed reaches $\approx c$.

\begin{figure}[H]
\begin{centering}
\includegraphics[width=0.35\paperwidth]{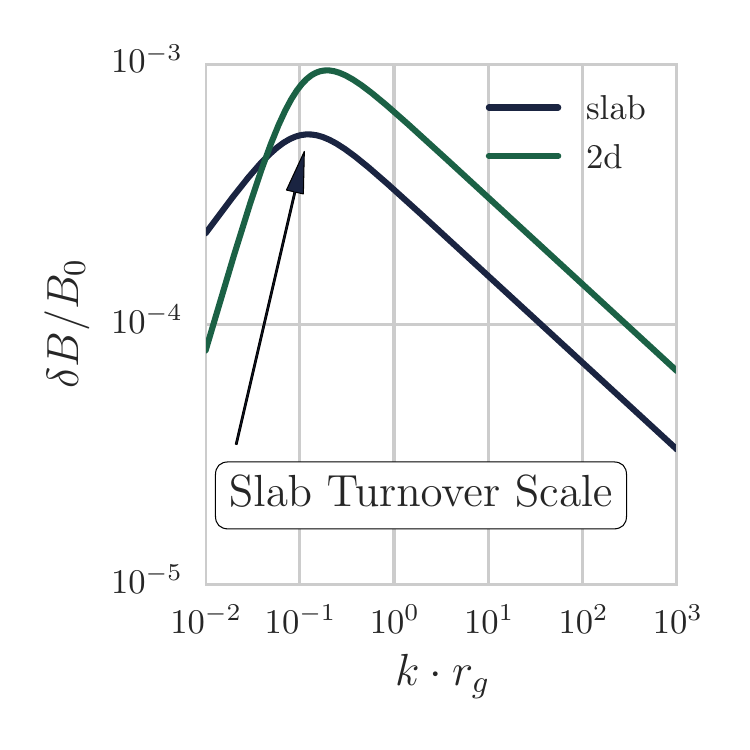}
\par\end{centering}
\centering{}\caption{\label{fig:wavespectrum}Example magnetic turbulence spectrum with
$s_{\mathrm{slab}}=5/3$, $q_{\mathrm{slab}}=0$, $s_{\mathrm{2d}}=5/3$,
$q_{\mathrm{2d}}=3/2$, $r_{\mathrm{slab}}=0.2$. The continuous spectrum
is approximated here by a sum over $2048$ discrete wave modes with
equal logarithmic spacing so that $\Delta\ln k=\Delta k/k$ is constant.}
\end{figure}

The spectrum of the waves is of the general form proposed by \citet{Shalchi2009a},
\begin{equation}
A_{k}^{2}\propto\Delta k\frac{k^{q}\ell^{q+1}}{\left(1+\left(k\ell\right)^{2}\right)^{\left(s+q\right)/2}}\label{eq:a2-spectrum}
\end{equation}
where $\Delta k$ is the spacing between waves and $\ell$ is the
turbulence turnover scale. The proportionality constant is chosen
so that the total turbulent wave energy is normalised to ${\delta B}^{2}$
from \eqref{dBaswaves}. Here $s$ and $q$ are dimensionless parameters
that shape the power law (see \figref{wavespectrum}) . This form
 smoothly interpolates between the two power law indices and therefore
obtains e.g. Kolmogorov and Goldreich-Sridhar turbulence as special
cases. We take the putative values of $s_{\mathrm{slab}}=5/3$, $q_{\mathrm{slab}}=0$,
$s_{\mathrm{2d}}=5/3$, $q_{\mathrm{2d}}=3/2$ as predicted by the
Kolmogorov spectrum \citep{Tautz2011,Hussein2015}. An example with
$r_{\mathrm{slab}}=0.2$ is seen in \figref{wavespectrum}.

\subsection{Numerical Setup}

\label{subsec:Numerical}

In the results presented in \secref{Results} below we chose the following
parameters (in simulation units where $r_{g}=2\pi$): wavenumbers
are uniformly distributed in log-space ($\Delta\ln k=\Delta k/k$
is constant) between the minimum and maximum $k_{min}=10^{-4}$ and
$k_{max}=10^{6}$ respectively. These values are chosen so as to allow
resonant interaction at most values of $\mu$. The spectral indices
are $s_{\text{slab}}=5/3$, $s_{\text{2d}}=5/3$, $q_{\text{slab}}=0$,
$q_{\text{2d}}=3/2$ \citep{Shalchi2009a}, and turbulence turnover
scales $\ell_{\text{slab}}=\ell_{\text{2d}}=1$. This assumption is
based on coupling between the ion gyroscale and turbulence length
scale from collisionless Landau damping, which results in turbulence
with turnover scale $\approx r_{g}$ \citep{Schekochihin2009}. Particles
are initially uniformly distributed in $\mu$-space ensuring they
interact with different parts of the spectrum. Their initial velocity
is chosen to be $v=0.1c$ (see discussion in \secref{Discussion}).
The number of waves and particles per seed is $n_{w}=4096$ and $n_{p}=256$
respectively. The number of random seeds corresponding to distinct
turbulence realisations for ensemble average $n_{s}=8$, which was
found to be enough to achieve convergence. The total run time is set
individually for each set of turbulence level, by using a small initial
run to determine the approximate value of the diffusion time $t_{s}$
(see \secref{Diffusion} and appendix \ref{sec:Calculation-of-Scattering}
below) and then running the full simulation with $t_{\mathrm{max}}\gtrsim100t_{s}$
in order to be able to capture the diffusion in each case. The integrator
used is \texttt{bulirsch-stoer} from \texttt{odeint} \citep{Ahnert2011}
with relative and absolute tolerance $\varepsilon_{\text{rel}}=\varepsilon_{\text{abs}}=10^{-9}$.
The particle trajectories are tracked and the scattering time $t_{s}$
and pitch-angle diffusion coefficient $D_{\mu\mu}$ are calculated.
As this can be done in more than one way we explain our calculation
method in \secref{Diffusion} below.

\section{Diffusion}

\label{sec:Diffusion}

In this section we motivate and explain our calculation of the pitch-angle
diffusion coefficient $D_{\mu\mu}$. We briefly review the significance
of this parameter and methods for measuring its value from simulation
results. It is seen that the choice of $D_{\mu\mu}$ is also reliant
on the choice of $\Delta t$, the time over which diffusion steps
are measured. Note that $\Delta t$ provides a timescale for Monte\textendash Carlo
simulations which make use of diffusion models discussed here, and
is distinct from the integration timestep of the simulation used in
this work (see \secref{Model}). We discuss choices of $\Delta t$
as they apply to numerical simulations and as they relate to the QLT
and Bohm approximations. 

\subsection{Definition of $D_{\mu\mu}$}

The particle population in a CR accelerator system can be encapsulated
in the multi-particle phase space distribution function $f\left(x^{i},p^{i},t\right)$
where $\int_{\Omega}f\,\mathrm{d^{3}x}\mathrm{d^{3}p}$ is the number
of particles in the phase-space volume $\Omega$, and $x^{i}$ and
$p^{i}$ are respectively spatial coordinates and momentum coordinates.
Its evolution is described by the Fokker-Planck equation \citep{Hall1967,Schlickeiser2002},
\begin{equation}
\partial_{t}f+v^{i}\partial_{x^{i}}f+p^{-2}\partial_{p^{i}}\left(p^{2}D^{p_{i}p_{j}}\partial_{p^{j}}f\right)=0\label{eq:badfokkerplanck}
\end{equation}
where $p$ is momentum and $D$ is the diffusion\emph{ }tensor\emph{.
}For a derivation and background on this equation see \citet{Hall1967,Newman1973,Urch1977,Schlickeiser2002}.
We use the convenient momentum-space basis $\left(p_{r},p_{\phi},p_{\mu}\right)$
of radial, gyrophase and pitch-angle cosine respectively. We assume
axisymmetry around $\boldsymbol{B}_{0}$, that scatterings are elastic,
and ignore perpendicular and gyrophase motion. This gives symmetry
in $x$, $y$, and $p_{\phi}$ and that $p_{r}=\left|p\right|$ is
constant, hence our remaining dependent variables are just $z$ and
$\mu$ this equation can be written as
\begin{equation}
\partial_{t}f+v\mu\partial_{z}f+\partial_{\mu}\left({D_{\mu\mu}}\partial_{\mu}f\right)=0\label{eq:mufokkerplanck}
\end{equation}
where by abuse of notation we have defined $D_{\mu\mu}$ as the component
of the tensor $D$ in the direction of $\mu$, i.e. $D^{p^{\mu}p^{\mu}}=D_{\mu\mu}$
and assumed all other components are zero.

The parameter $D_{\mu\mu}$ typically depends on the pitch angle as
well as the turbulence details and determines the particle trajectory
by encapsulating the details of the turbulent wave-particle interaction
(we neglect particle-particle interation). The problem of defining
and measuring the diffusion coefficient $D_{\mu\mu}$ has been discussed
at length in the literature, both in terms of the appropriate timescale
over which to measure \citep{Giacalone1999,Shalchi2006,Spanier2011}
and the appropriate way to calculate it  \citep{Knight2011,Shalchi2011}.
The difficulty arises from the fact that, unlike the classical hard-sphere
scattering case, the diffusion does not occur in response to discrete
events but rather a gradual collective interaction with the whole
spectrum of magnetic turbulence simultaneously.

While there are several possible prescriptions for calculating $D_{\mu\mu}$,
for the purposes of this work the mean square deviation (MSD) form
\citep{Jokipii1966,Blandford1987,Tautz2013} is used,
\begin{equation}
D_{\mu\mu,\text{MSD}}\equiv\left<\frac{\left(\Delta\mu\right)^{2}}{\Delta t}\right>\label{eq:msd1}
\end{equation}
where $\Delta\mu=\mu\left(t\right)-\mu\left(0\right)$ and the chevrons
indicate an average over ensemble\footnote{If the turbulence is ergodic, as is commonly assumed \citep{Lemoine2001},
then this is equivalent to a time average. }. This form is suitable because it allows the parameter $\Delta t$
to be tuned, and as we will discuss in \subsecref{Time-Scales} this
determines what timescales can be resolved. 

The MSD form is the most straightforward way of calculating $D_{\mu\mu}$.
We note, however, that other methods exist. Of particular interest
are the Taylor-Green-Kubo (TGK) integral and derivative methods. The
TGK integral form \citep{Tautz2013}, $D_{\mu\mu,\text{TGKI}}\equiv\int_{0}^{\infty}\mathrm{d}t\left\langle \dot{\mu}\left(t\right)\dot{\mu}\left(0\right)\right\rangle $
where overdot indicates time derivative, is more amenable than MSD
to analytic work. However this quantity is unsuitable for numerical
work as the integral does not converge when the upper limit is taken
to infinity. In numerical approximations, when the upper integration
limit is taken to be $\Delta t$, it is identical to the MSD form
\citep{Shalchi2011}. A third method, the TGK derivative form \citep{Tautz2013}
$D_{\mu\mu,\text{TGKD}}\equiv\frac{\mathrm{d}}{\mathrm{d}t}\left\langle \left(\Delta\mu\right)^{2}\right\rangle $,
is the limit of $D_{\mu\mu,\text{MSD}}$ as $\Delta t\rightarrow0$,
however this form cannot be used since for $\Delta t$ too short only
ballistic motion will be seen\footnote{For time scales much shorter than the wave crossing time, $t_{w}=2\pi/\left(\mu vk_{\mathrm{max}}\right)$
the local $B$-field is roughly constant and the particle exhibits
unperturbed gyromotion.} \citep{Zank2014}.

\subsection{On the Proper Choice of the Diffusion Timestep $\Delta t$\label{subsec:Time-Scales}}

 As we show below in \secref{Results}, the value of $D_{\mu\mu,\mathrm{MSD}}$
in \eqref{msd1} is highly sensitive to the choice of $\Delta t$,
the time over which $\Delta\mu$ is measured. In order to obtain a
physically meaningful value for $D_{\mu\mu}$ the following must be
considered. In the limit $\Delta t\rightarrow0$ the diffusion coefficient
approaches zero, regardless of the details of the diffusion, because
the numerator in \eqref{msd1} is second order in $\Delta t$, while
the denominator is only first-order. On the other hand the value of
$\Delta t$ cannot be too long. Since $\Delta\mu$ can be at most
$2$ an arbitrarily large value of $\Delta t$ causes $D_{{\mu\mu}}$
to vanish.

Two useful timescales which will be employed are the scattering time
$t_{s}$ and the diffusion time $t_{D}$. The scattering time $t_{s}$
can be defined as the expected time for the particle's pitch angle
to change by a $\Delta\vartheta$, typically $90\textdegree$, or
equivalently, as in this paper, as the decorrelation time 
\begin{equation}
t_{s}=\left\langle \int_{0}^{\infty}C_{\mu}\left(t,\tau\right)\mathrm{d}t\right\rangle _{\tau}\label{eq:definition-of-ts}
\end{equation}
 \citep{Casse2002}, where $C_{\mu}\left(t,\tau\right)=\left\langle \mu\left(t+\tau\right)\mu\left(t\right)\right\rangle /\left\langle \mu\left(t\right)^{2}\right\rangle $
is the autocorrelation of $\mu$ at time $t$ and $\left\langle \cdot\right\rangle _{t}$
indicates a time average. See appendix \ref{sec:Calculation-of-Scattering}
for further details. We similarly define the diffusion time $t_{D}=1/\left\langle D_{\mu\mu}\right\rangle $,
i.e. the time taken for the particle to significantly change pitch-angle
so that $\Delta\mu\approx1$.

The assumption that particles interact only resonantly with waves
(as is assumed in QLT) requires that differences in $\mu$ are measured
over many gyrotimes, so that the force contributions from nonresonant
waves average to zero. In the weak turbulence regime QLT is known
to be a good approximation \citep{Shalchi2009c}, and so we retain
this constraint in order to recover QLT in this limit. 

The Bohm approximation, on the other hand, assumes that the scattering
time is roughly equal to the gyrotime, so that $\Delta t$ may not
be much less than $t_{g}$. However as long as $\Delta\mu\ll1$ values
of $\Delta t>t_{g}$ may be used.

Neither the Bohm nor QLT type models include a description of frequent
scattering, i.e. more than once per gyrotime. Since the particle may
scatter to a significantly different pitch angle within a single gyrotime
it can have a different resonant wavelength and hence interact with
a different portion of the turbulence spectrum. Moreover at this point
the particle is no longer undergoing gyromotion, and cannot be treated
as a ``scattering gyrocentre''. We can estimate the turbulence strength
at which the scattering becomes more frequent than the gyration by
equating the gyrotime $t_{g}$ with the diffusion time $t_{D}$. In
order to have at least one scattering per diffusive timestep in this
case we then must have $t_{g}\leq\Delta t$. However it is found that
for strong turbulence, the scattering time is on the order of, or
shorter than the gyrotime.

\begin{center}
\par\end{center}

To conclude, any timestep must satisfy several upper and lower limits.
Compatibility with the assumptions of QLT requires that $t_{g}\ll\Delta t\ll t_{D}$,
and Bohm requires $t_{g}\lesssim\Delta t\ll t_{D}$. This imples that
in order to use a diffusion model we must have $t_{g}<t_{D}$. We
discuss below the conditions under which this requirement is met.

\subsection{Anomalous Diffusion}

Classical diffusion processes resulting from discrete scattering events
in unbounded regions (e.g. gas diffusion) exhibit displacements of
the form $\left\langle \left(\Delta x\right)^{2}\right\rangle \propto\Delta t$
for all timesteps $\Delta t$ much greater than the scattering time,
and so $D_{xx}=\frac{\left\langle \left(\Delta x\right)^{2}\right\rangle }{\Delta t}$
is independent of $\Delta t$. This is also the case for the Bohm
and QLT models. However, when turbulence is so strong that within
each timestep $\mu$ changes significantly, this behaviour is not
observed and the time dependence of the MSD is nonlinear \citep{Metzler2000}.
Hence we consider generalised diffusion models where 
\begin{align}
\left\langle \left(\Delta\mu\right)^{2}\right\rangle  & \propto\Delta t^{b}\label{eq:anomalous_b}
\end{align}
 with $b\neq1$. This is known as \emph{anomalous diffusion} (see
e.g. \citet{Pommois2007,Shalchi2009a,Bykov2017}).  The physical
mechanisms by which $D_{{\mu\mu}}$ can depend nonlinearly on $\Delta t$
are: directly though time-dependent turbulence, and indirectly through
dependence on $\mu$. Solving \eqref{anomalous_b} for $b$ gives
$b=\frac{\mathrm{d}\ln D_{\mu\mu}}{\mathrm{d}\ln\Delta t}+1$ and
this is used to identify the diffusion regime, as in \Cref{fig:time-test,fig:time-test-sr1.0}.
The turbulence is time-independent since wave propagation and feedback
are not included so any anomalous diffusion must be inherent to the
scattering process. When the diffusion is not anomalous the choice
of $\Delta t$ is seen to be arbitrary (within the constraints of
\subsecref{Time-Scales}) and in practice is chosen to be $20t_{g}$.
Notably in the case of bounded diffusion like that of $\mu$, there
is always a non-physical $b=-1$ regime for sufficiently large $\Delta t$.

\section{Results}

\label{sec:Results}

In this section we show the results of our simulations or particle
transport. From the gathered data we calculate $D_{\mu\mu}$, and
its time-dependence. We separately calculate the scattering time $t_{s}$
and finally use this data to find the Bohm exponent $\alpha$ and
give its dependence on the turbulence strength.

\subsection{Choice of $\Delta t$}

\label{subsec:choiceofdeltat}

\begin{figure*}
\begin{centering}
\emph{\includegraphics[width=0.33\textwidth]{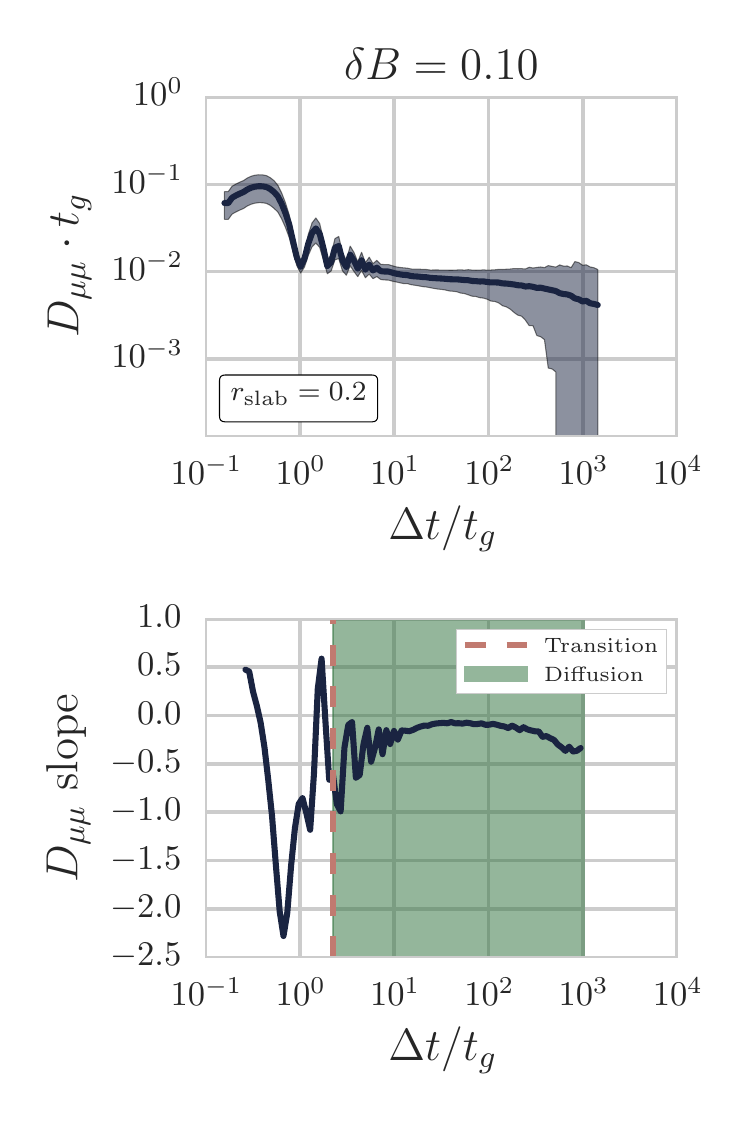}\includegraphics[width=0.33\textwidth]{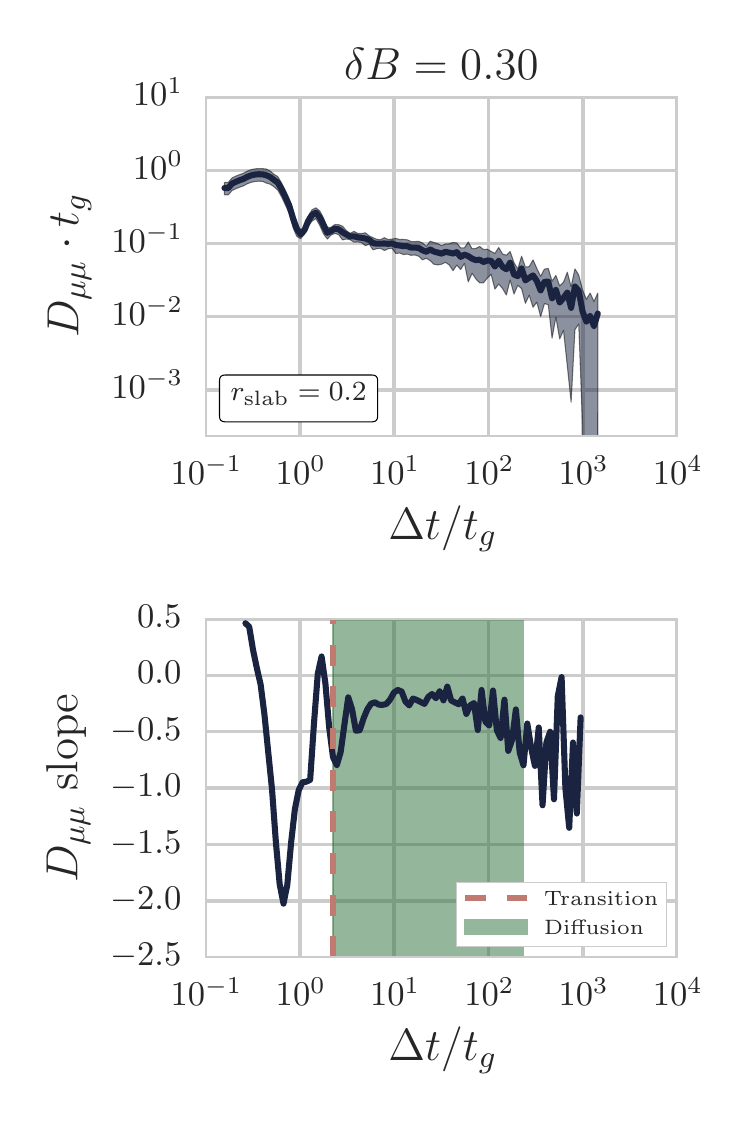}\includegraphics[width=0.33\textwidth]{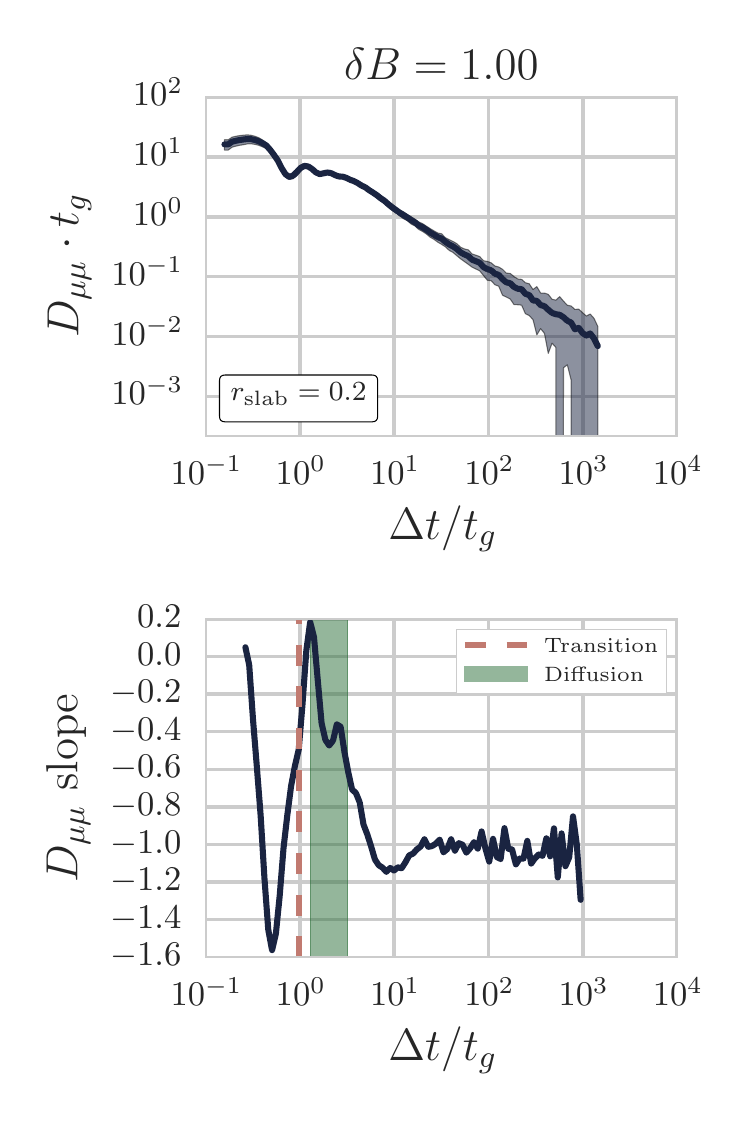}}
\par\end{centering}
\begin{centering}
\par\end{centering}
\centering{}\caption{\label{fig:time-test}The upper plots show $D_{{\mu\mu}}$ (pitch-angle
averaged) as a function of $\Delta t$ for $r_{\mathrm{slab}}=0.2$,
with the solid line for the particle average and the shaded area representing
one standard deviation. \label{fig:time-test-2}The lower plots show
the average slope of $D_{{\mu\mu}}$ as a function of $\Delta t$,
highlighting the transition from ballistic to subdiffusive behaviour
($\frac{\mathrm{d}\ln D_{\mu\mu}}{\mathrm{d}\ln\Delta t}<0$, dashed
red line), and the diffusive regime ($\frac{\mathrm{d}\ln D_{\mu\mu}}{\mathrm{d}\ln\Delta t}\approx0$,
green shading). Notice that the diffusive range shrinks as the turbulence
strength increases. These plots are in simulation units where $B_{0}=1$.}
\end{figure*}

\begin{figure*}
\begin{centering}
\emph{\includegraphics[width=0.33\textwidth]{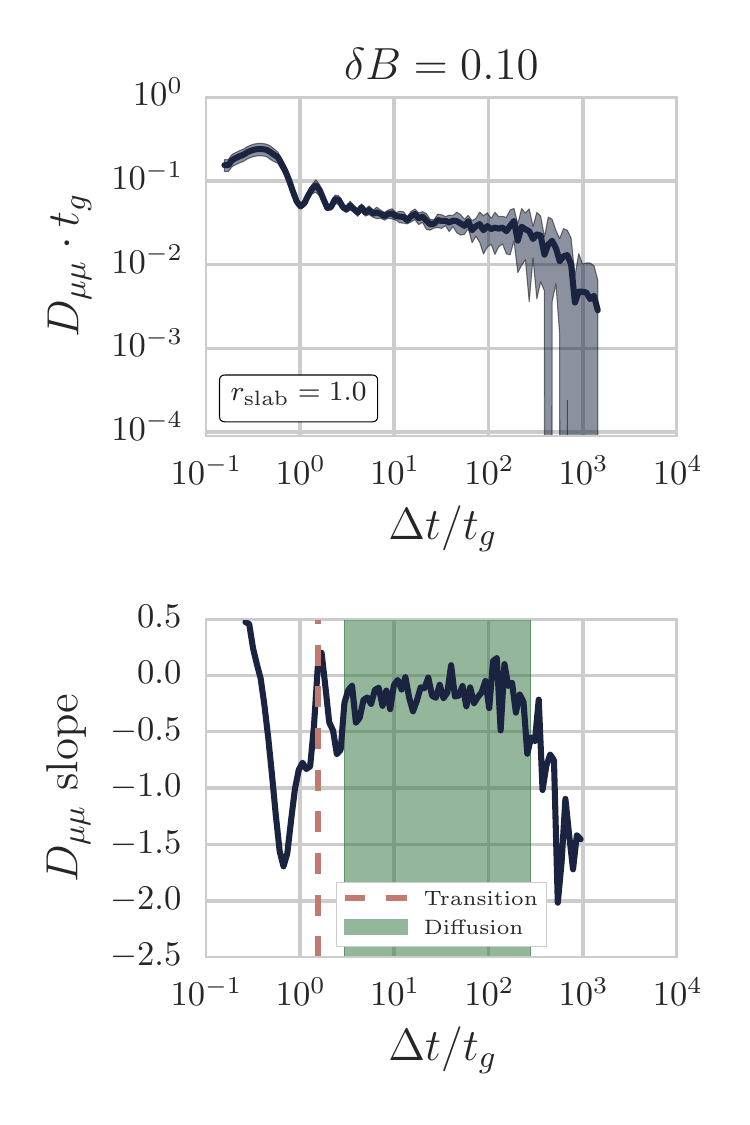}\includegraphics[width=0.33\textwidth]{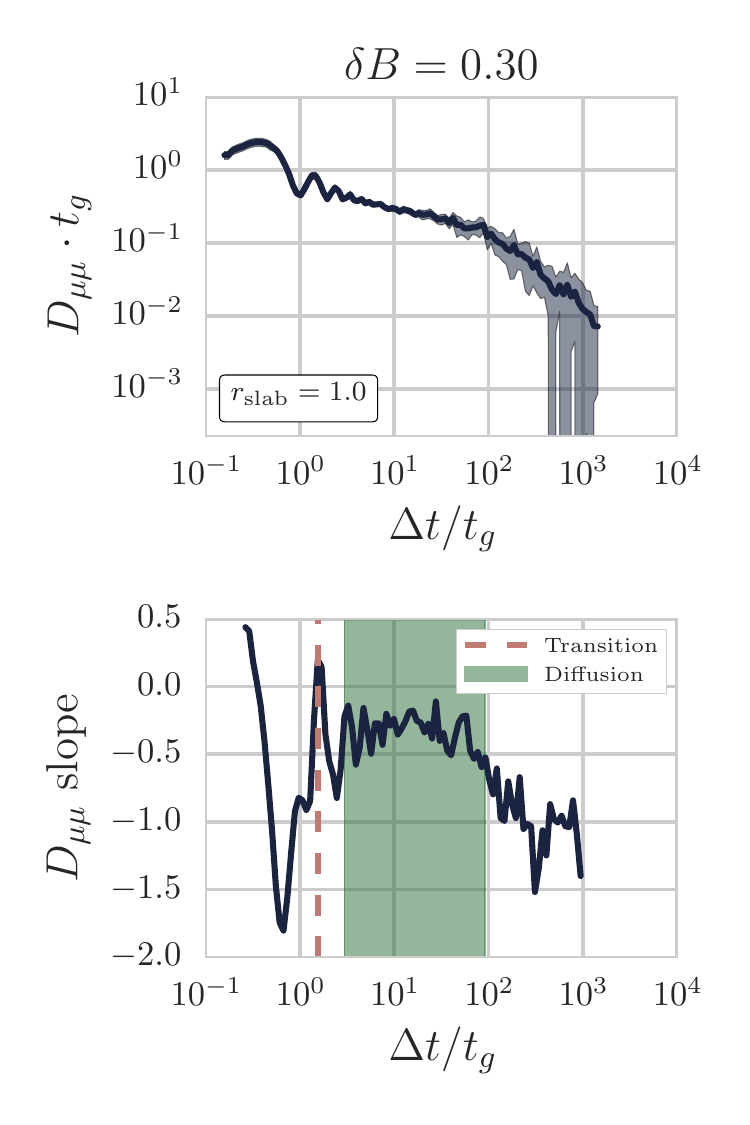}\includegraphics[width=0.33\textwidth]{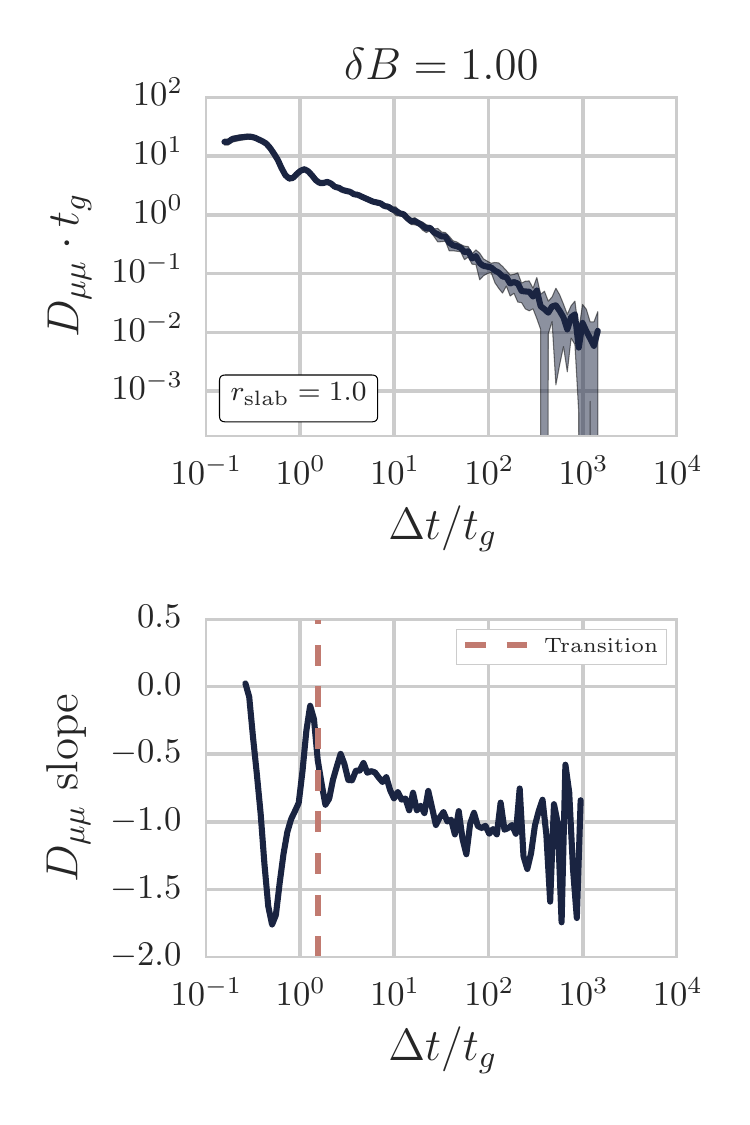}}
\par\end{centering}
\begin{centering}
\par\end{centering}
\begin{centering}
\par\end{centering}
\centering{}\caption{\label{fig:time-test-sr1.0}Same as \figref{time-test}, but for $r_{\mathrm{slab}}=1$
(pure slab turbulence). The $\delta B/B_{0}=1$ case has no significant
time region in which behaviour is diffusive since $t_{D}<t_{g}$ and
the middle term of \eqref{dmmucases} vanishes.}
\end{figure*}

In \Figref{time-test,time-test-sr1.0} (upper panels) we plot $D_{{\mu\mu}}$
against $\Delta t$ for various turbulence levels. The upper panels
show $D_{\mu\mu}$ as a function of $\Delta t$. The main features
are the initial ballistic phase, the diffusive phase ($b\approx1$),
and the subdiffusive (\textbf{$b<1$}) tail. As the turbulence level
is increased the diffusive region shrinks and eventually disappears
entirely. \Figref{time-test,time-test-sr1.0} (lower panels) plot
the slope $\frac{\mathrm{d}\ln D_{\mu\mu}}{\mathrm{d}\ln\Delta t}$
and allow us to clearly delineate the end of the ballistic phase (red
dashed line) and the diffusive phase (green shading). This shows that
in the weak turbulence case ($\delta B/B_{0}=0.1$) the behaviour
after several gyrotimes become approximately diffusive over a wide
range of $\Delta t$, as predicted by QLT. 

We find that $D_{\mu\mu}$ initially increases quadratically while
$\Delta t$ is short enough that the particle motion is ballistic
\citep{Tautz2011,Tautz2013}. It then peaks, may remain constant for
some range of of $\Delta t$ (depending on turbulence level), and
then diminishes linearly due to the boundedness of $\mu$. Hence the
diffusion coefficient is approximately described by 
\begin{equation}
D_{\mu\mu}\left(\Delta t\right)\propto\begin{cases}
\Delta t^{2} & \Delta t<t_{g}\\
\Delta t^{0} & t_{g}<\Delta t<t_{D}\\
\Delta t^{-1} & \Delta t>t_{D}
\end{cases}\label{eq:dmmucases}
\end{equation}
corresponding to the cases of anomalous diffusion index $b=$$3$,
$1$, and $0$ respectively. There is also a sinusoidal component
due to the gyromotion of the particle, which causes the observed turbulent
magnetic field to rotate at the angular gyrofrequency $\omega_{g}=2\pi/t_{g}$. 

The difference in $r_{\text{slab}}$ and hence lower two-dimensional
turbulence strength in \figref{time-test-sr1.0} has the effect of
enhancing the diffusion by a factor of order unity..

\begin{figure}
\begin{centering}
\includegraphics[width=0.5\columnwidth]{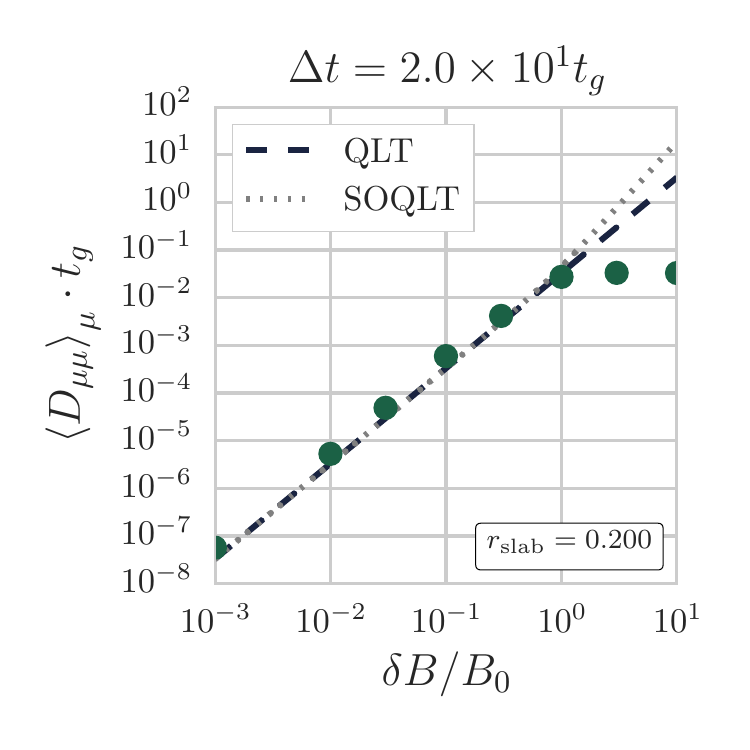}\includegraphics[width=0.5\columnwidth]{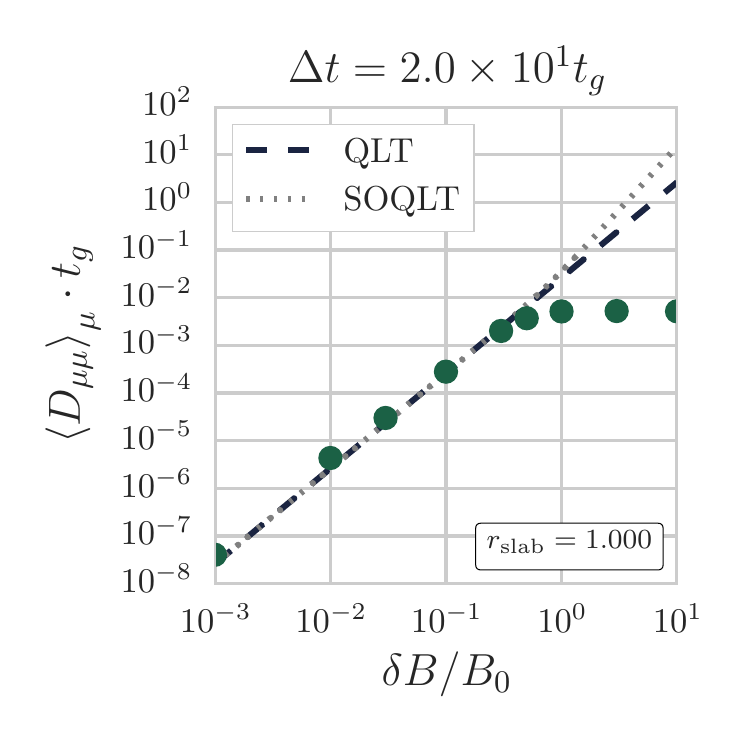}
\par\end{centering}
\caption{$D_{\mu\mu}$ (pitch-angle averaged) measured at $\Delta t=20t_{g}$
for various (slab-only) turbulence levels. Good agreement with the
(SO)QLT models for weak turbulence. For turbulence levels $\delta B\gtrsim1$
the measurement time $\Delta t$ of $20t_{g}$ is no longer within
the diffusion regime (see \figref{time-test}) and so the values are
no longer meaningful.\label{fig:-measured-at20tg}}
\end{figure}

In \figref{-measured-at20tg} we show the diffusion coefficient as
a function of turbulence level. It initially increases as $\delta B^{2}$
(QLT regime), gradually flattens around $\delta B/B_{0}=10^{-1}$
and stays roughly constant thereafter. We argue that this flattening
is not physical but an artifact of the fact the we are measuring a
bounded quantity $\Delta\mu$ over a time period longer than its dynamical
time $1/D_{\mu\mu}$. Indeed, from \figref{time-test} it can be seen
that only for low-turbulence cases is there a region where 
\begin{equation}
\frac{\mathrm{d}D_{\mu\mu}}{d\Delta t}\approx0\label{eq:dmumu_is_flat}
\end{equation}
(or equivalently that $b=1$) and the behaviour can be considered
classically diffusive (i.e. independent of $\Delta t$). Once this
region vanishes (see \figref{time-test}) the process can no longer
be treated as non-anomalous diffusion (c.f. \citet{Qin2009}). 

\subsection{\label{subsec:Validity-of-Bohm}Validity of Bohm Approximation}

\begin{figure}
\begin{centering}
\emph{\includegraphics[width=0.5\columnwidth]{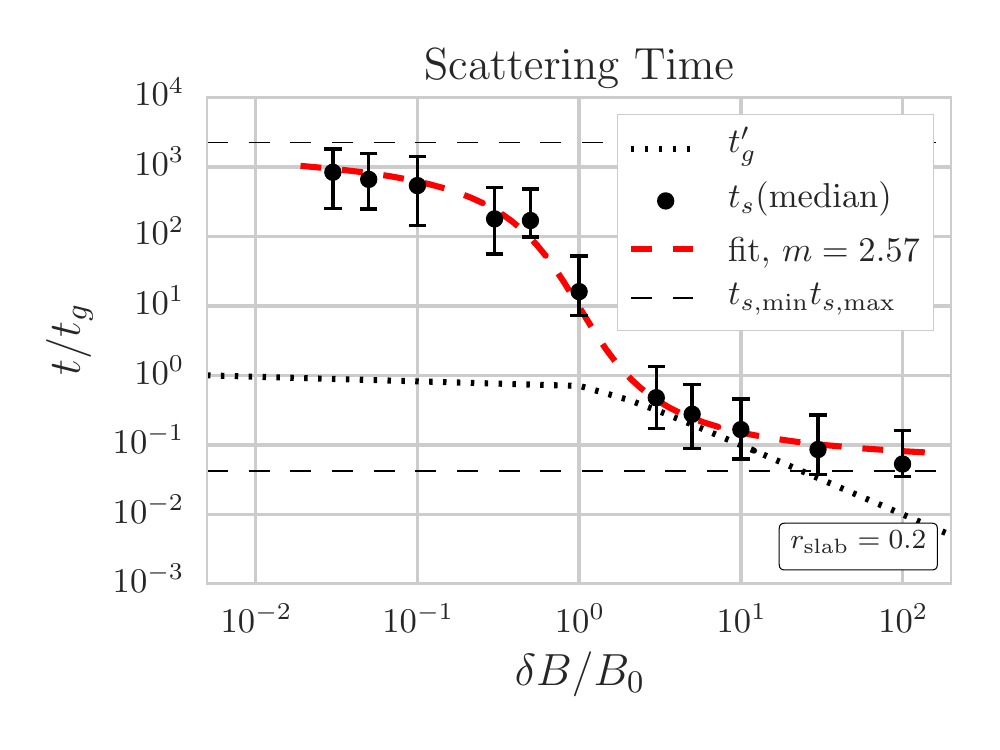}}
\par\end{centering}
\begin{centering}
\emph{\includegraphics[width=0.5\columnwidth]{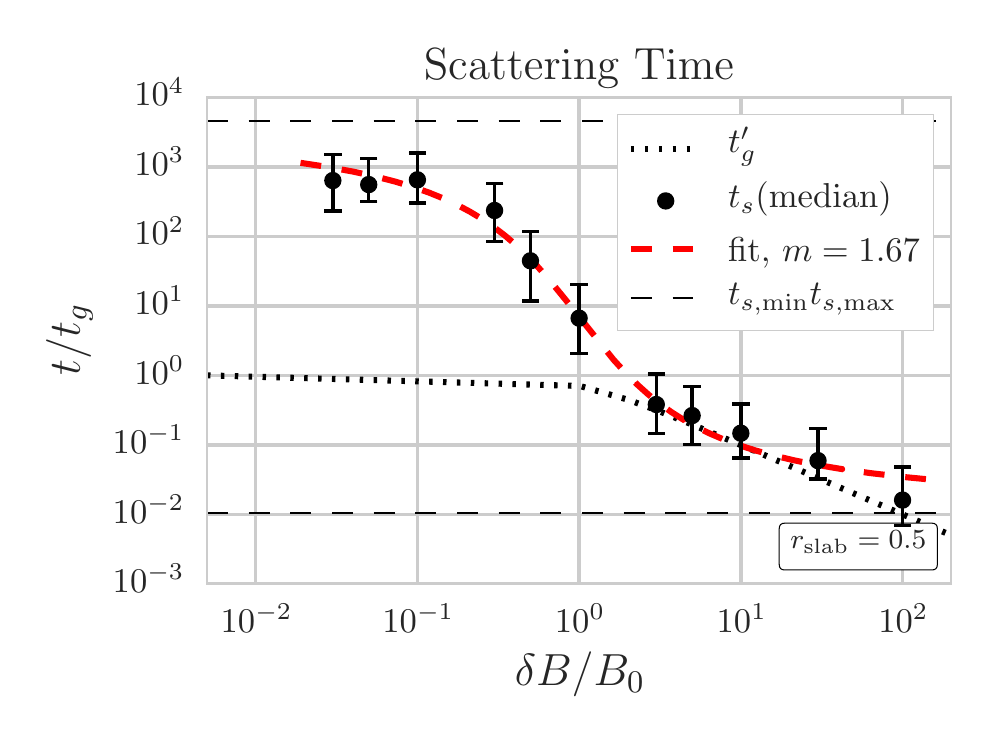}}
\par\end{centering}
\begin{centering}
\emph{\includegraphics[width=0.5\columnwidth]{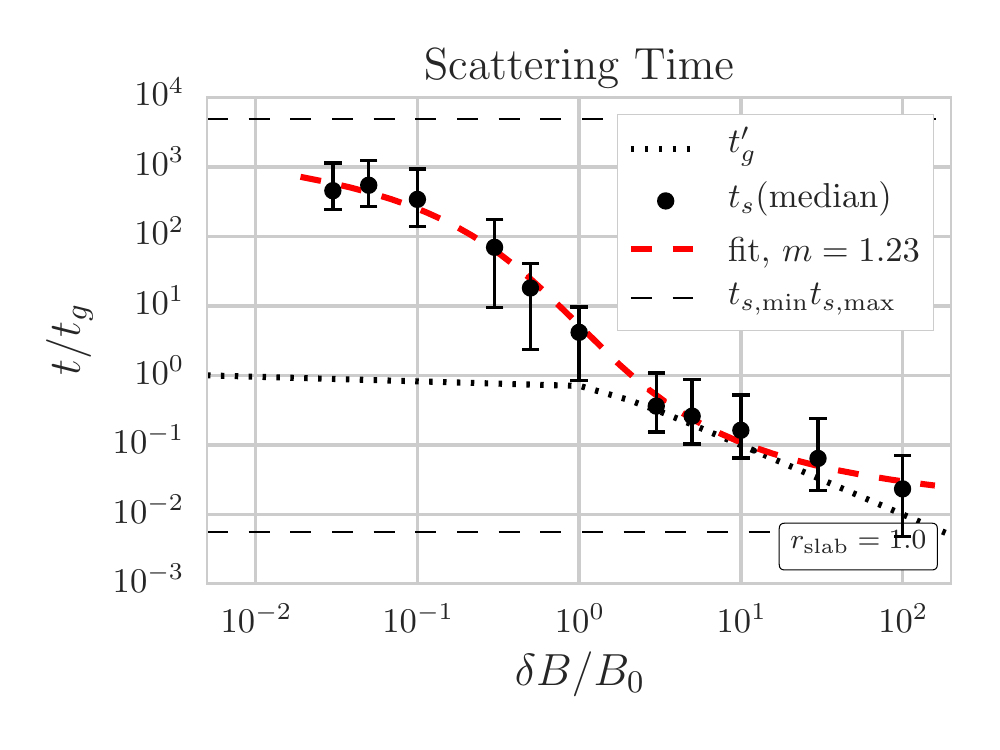}}
\par\end{centering}
\centering{}\caption{Scattering time $t_{s}$ as a function of turbulence level $\delta B/B_{0}$
with $r_{slab}=0.2$, $0.5$ and $1.0$. Here $t_{g}^{\prime}$ represents
the ``effective'' gyrotime determined by the effective field $B_{\text{eff}}$.
Points represent the ensemble and pitch-angle median of the measured
scattering times for each turbulence level, and the error bars the
first and third quartile . The red line is a fit with \eqref{fitted_model},
with the parameters indicated on the respective graph and in \tabref{Best-fit-parameters}.
 \label{fig:Scattering-time-as}}
\end{figure}

One can model the diffusion of charged particles as a power law relationship
between mean free path $\lambda_{\mathrm{mfp}}$ and  gyroradius
\citep{Pommois2007}. Here $\lambda_{\mathrm{mfp}}$ refers to the
expected value of the distance travelled by a particle in the time
it takes for $\vartheta$ to change by $\pi/2$; this corresponds
to the mean free path between scatterings in the case of only right-angle
collisions \citep{Ellison1990,Summerlin2011},
\begin{equation}
\lambda_{\mathrm{mfp}}=\eta r_{g}^{\alpha}\label{eq:bohm-type}
\end{equation}
where $\eta$ is the Bohm coupling constant and $r_{g}$ is suitably
normalised. In particular the physical role of $\eta$ is to represent
the effectiveness of the magnetic turbulence in diffusing the particles
(see also \subsecref{diffusionmodels}). We refer to these as \emph{Bohm-type}
models. It is intentionally not specified whether the gyration radius
$r_{g}$ refers to the radius of the gyration caused by the background
field $B_{0}$, the total effective field $B_{\mathrm{eff}}=\sqrt{B_{0}^{2}+\delta B^{2}}$
or an intermediate approximation, as different authors make different
choices here (see \citet{Vladimirov06}). In this work the ``effective''
field gyroradius and gyrotime are denoted $r_{g}^{\prime}$ and $t_{g}^{\prime}$
respectively, as in \figref{Scattering-time-as}.

In this work we take the ``Bohm approximation'' to mean $\alpha\approx1$.
This can be formulated equivalently as 
\begin{equation}
t_{s}\approx t_{g}\label{eq:tseqtg}
\end{equation}
(as long as $\eta$ is of order unity) meaning scattering occurs once
per gyrotime. We present the ratio of the scattering time to gyrotime
in \figref{Scattering-time-as}. From the figure we see that this
form of Bohm approximation is valid only around $\delta B/B_{0}\approx1$
for the unmodified gyrotime $t_{g}$. However we find that the modified
form of the Bohm approximation, $t_{g}^{\prime}\approx t_{s}$, is
valid until $\delta B/B_{0}\approx10$ . This behaviour does not
continue to higher turbulence levels, and the scattering time instead
asymptotes. Running the simulation with a smaller $\left[k_{\mathrm{min}},k_{\mathrm{max}}\right]$
range confirms that this behaviour is due to the finite wavelength
cutoff in the turbulence spectrum. 

Heuristically we can fit this with a sigmoid function, in \figref{Scattering-time-as}
we take
\begin{equation}
\log_{10}t_{s}/t_{g}=t_{s,\mathrm{min}}+\frac{t_{s,\mathrm{max}}-t_{s,\mathrm{min}}}{2}\left(1+\frac{m}{\pi}\arctan(-m\log_{10}\left(\delta B/B_{0}\right)\right)\label{eq:fitted_model}
\end{equation}

where parameters $t_{s,\mathrm{max}}$ and $t_{s,\mathrm{min}}$ are
respectively the low and high turbulence limits of the scattering
time, and $m$ (not to be confused with the particle mass) is a free
parameter determining the slope of the transition region. The best
fit parameters from our results are as given in \tabref{Best-fit-parameters}.
\begin{center}
\begin{table}[H]
\begin{centering}
\begin{tabular}{llccc}
\toprule 
$r_{\mathrm{slab}}$ &  & $t_{s,\mathrm{max}}/t_{g}$ & $t_{s,\mathrm{min}}/t_{g}$ & $m$\tabularnewline
\midrule
\midrule 
$0.2$ &  & $2.228\times10^{3}$ & $4.178\times10^{-2}$ & $2.573$\tabularnewline
$0.5$ &  & $4.571\times10^{3}$ & $1.038\times10^{-2}$ & $1.668$\tabularnewline
$1.0$ &  & $4.898\times10^{3}$ & $5.546\times10^{-3}$ & $1.230$\tabularnewline
\bottomrule
\end{tabular}
\par\end{centering}
\caption{\label{tab:Best-fit-parameters}Best fit parameters for \eqref{fitted_model}
applied to the data in \figref{Scattering-time-as}. Parameters $t_{s,\mathrm{max}}$
and $t_{s,\mathrm{min}}$ are respectively the low and high turbulence
limits of the scattering time, and $m$ determines the slope of the
transition region.}
\end{table}
\par\end{center}

 Since $\lambda_{\mathrm{mfp}}=\eta r_{g}^{\alpha}$, $r_{g}=\frac{\gamma mv}{eB}\left(1-\mu^{2}\right)^{1/2}$,
and $t_{s}=\frac{\lambda_{\mathrm{mfp}}}{v}$, the following relation
holds:
\begin{equation}
t_{s}\propto v^{\alpha-1}B^{-\alpha}.\label{eq:tspropbalpha}
\end{equation}
It is clear from \figref{Scattering-time-as} and \eqref{tspropbalpha}
that the value of $\alpha$ must vary significantly between turbulence
regimes, since otherwise a single power law would be observed. Hence
no model with constant $\alpha$ can cover the entire turbulence range.
\citet{Vladimirov2009} implicitly recognises this when interpolating
between different values of $\alpha$ for different particle energies.

\begin{figure}
\begin{centering}
\emph{\includegraphics[width=0.5\columnwidth]{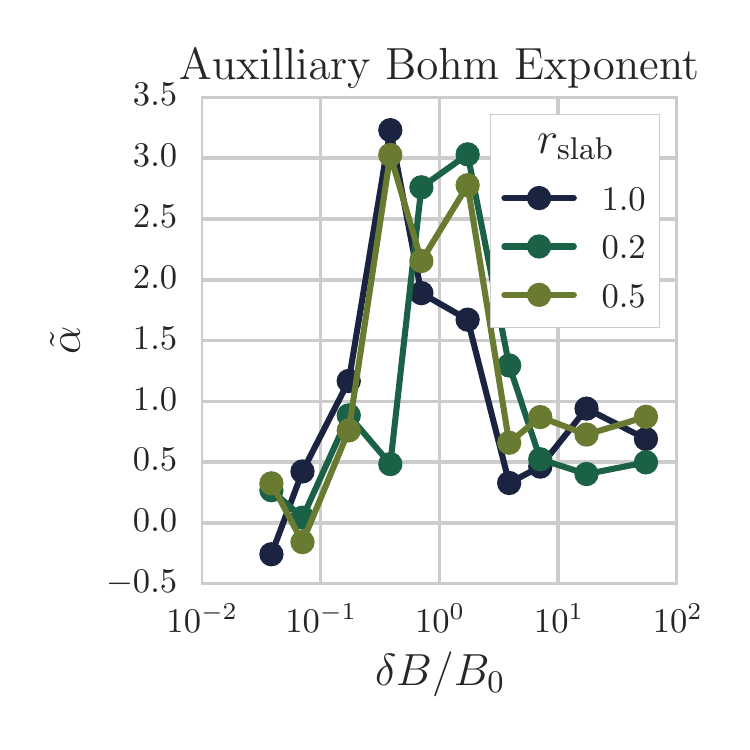}}
\par\end{centering}
\centering{}\caption{\label{fig:alpha-bohm-expo}Auxillary Bohm exponent as a function
of turbulence level for each of the considered values of $r_{slab}$
. This is calculated from the data in (\ref{fig:Scattering-time-as}).
It shows how a Bohm-type model can interpolate from weak up to intermediate
and strong turbulence, by choosing the approximate values $\tilde{\alpha}=0$,
$2.5$ and $0.7$ for these regions respectively.}
\end{figure}

From \eqref{tspropbalpha} the Bohm exponent $\alpha$ can be expressed
as 
\[
\alpha=-\frac{\mathrm{d}\ln t_{s}}{\mathrm{d}\ln B_{\mathrm{eff}}}.
\]
Reformulating in order to isolate the dependence on $\delta B$, 
\begin{align*}
\alpha & =-\frac{\mathrm{d}\ln t_{s}}{\mathrm{d}\ln\delta B}\frac{\mathrm{d}\ln\delta B}{\mathrm{d}\ln B_{\mathrm{eff}}}\\
 & =-\frac{\mathrm{d}\ln t_{s}}{\mathrm{d}\ln\delta B}\frac{B_{\mathrm{eff}}^{2}}{{\delta B}^{2}}\\
 & =\tilde{\alpha}\left(1+\left(\frac{\delta B}{B_{0}}\right)^{-2}\right)
\end{align*}
where $\tilde{\alpha}=-\frac{\mathrm{d}\ln t_{s}}{\mathrm{d}\ln\delta B}$
is the slope of the data presented in \figref{Scattering-time-as}.
The values of this ``auxillary Bohm exponent'' are calculated using
a simple finite difference method and are plotted in \figref{alpha-bohm-expo}.
These show that $\alpha$ varies over the intermediate turbulence
regime, and indeed that for $r_{\mathrm{slab}}=1$ and $\delta B/B_{0}\gtrsim1$
we find $\tilde{\alpha}\approx1$, in good agreement with the ``Bohm
approximation'' that $\alpha=1$. Using these numerical results as
model for diffusive motion, a Monte-Carlo simulation of particle acceleration
can accurately account for the turbulent dependence of mean free path
in the entire regime of turbulence strength using values of $\alpha$
presented here. In particular the region between $0$ and $\approx10$,
which as we have discussed is not covered by existing models, is covered.

We can estimate the value of the parameter $\eta$ in the regions
where the slope of $t_{s}$ and $t_{g}^{\prime}$ are similar and
hence $\eta\approx t_{s}/t_{g}^{\prime}$. For weak turbulence ($\delta B/B_{0}\ll1$),
$\alpha\approx0$ and we find $\eta\approx10^{3}$. For strong turbulence
($\delta B/B_{0}\gg1$) we find $\eta\approx1$. This is in agreement
with the classical Bohm heuristic of scattering ``once per gyrotime''. 

\section{Discussion}

\label{sec:Discussion}

\subsection{Diffusion Models}

\label{subsec:diffusionmodels}

Bohm-type models supported by numerical evidence of 1-D Monte-Carlo
simulations with power law Alfvén spectrum and observational evidence
of ISEE-3 in the solar wind \citep{Giacalone1992}, and more recently
by PIC simulations when self-generated turbulence is included \citet{Caprioli2014}.
It is not clear what the appropriate value of $\alpha$ is in each
case, but some models are given in \tabref{Spectral-Indices}.
\begin{center}
\begin{table}[H]
\begin{centering}
\begin{tabular}{ll}
\toprule 
Model & $\alpha$\tabularnewline
\midrule
\midrule 
Constant/Low Energy\textsuperscript{\ref{enu:ell}} & $0$\tabularnewline
\midrule
Kolmogorov\textsuperscript{\ref{enu:amat}} & $1/3$\tabularnewline
\midrule
Kraichnan\textsuperscript{\ref{enu:amat}} & $1/2$\tabularnewline
\midrule
Galactic\textsuperscript{\ref{enu:Shalch}} & $0.6\pm0.1$\tabularnewline
\midrule
Bohm\textsuperscript{\ref{enu:barsum}} & $1$\tabularnewline
\midrule
High-Energy\textsuperscript{\ref{enu:topty}} & $2$\tabularnewline
\bottomrule
\end{tabular}
\par\end{centering}
\begin{inparaenum}[\expandafter \textsuperscript a]
\item \citet{Ellison1990}\label{enu:ell}
\item \citet{Amato2005}\label{enu:amat}
\item \citet{Shalchi2009c}\label{enu:Shalch}
\item \citet{Baring2009,Summerlin2011}\label{enu:barsum}
\item \citet{Toptygin1985}\label{enu:topty} 
\end{inparaenum}

\caption{\label{tab:Spectral-Indices}Values for the parameter $\alpha$ from
various models, where all but the Galactic value are based on theory.
Our model gives a value of $\alpha$ which varies with turbulence
strength, but includes each of these at particular values of $\delta B/B_{0}.$}
\end{table}
\par\end{center}

As discussed in \citet{Giacalone1992} a physical justification for
the value of $\alpha$ is difficult due to the fact that ``scattering
events'' are a simplification of the model and not an actual physical
phenomenon. On the other hand \citet{Summerlin2011} argue that $\alpha=1$
(the ``Bohm limit'') is necessary for ``physically meaningful diffusion''
while claiming results are not strongly sensitive to this value. It
can be shown \citep{Shalchi2009b} that if the diffusion time is
much shorter than the gyrotime then this assumption is equivalent
to the claim that $D_{\mu\mu}\propto\frac{\delta B}{B_{0}}$, i.e.
the diffusion is linearly dependent on the turbulence, as opposed
to the quadratic dependence of QLT \citep{Shalchi2015}. Heuristically
one can see this by noting that $\lambda_{\mathrm{mfp}}\propto D_{\mu\mu}^{-1}$
and $r_{g}\propto{\delta B}^{-1}$. In the limits of low and high
energy particles where $r_{g}<1/k_{\mathrm{max}}$ or $r_{g}>1/k_{\mathrm{min}}$,
there exists no wave with $k\approx1/r_{g}$ and so resonant interaction
is impossible and one obtains $\alpha=0$ and $\alpha=2$ respectively
\citep{Vladimirov2009}. 

Choosing a value of $\eta$, the Bohm coupling constant, is difficult,
since it is not predicted by the model. It is a significant parameter,
since it governs the anisotropy of the scattering \citep{Giacalone1999}
as well as the rate of energy gain in DSA \citep{Dermer2009}. It
is suggested, for example, by \citet{Vladimirov2009}, that $\eta$
may correspond to the magnetic correlation length for low energy ($\alpha=0$)
particles. Comparison of Monte-Carlo and PIC results for laboratory
plasmas found good agreement if a value of $\eta$ between $6$ and
$25$ is used \citep{Bultinck2010}, while observations of Cas A
suggest values between $2$ and $36$ \citep{Stage2006}. As shown
in \figref{Scattering-time-as} our results are consistent with $\eta\approx1$
in the region of $\delta B/B_{0}\approx3$, since here $t_{s}\approx t_{g}^{\prime}$.

Several other works have used numerical methods to find diffusion
coefficients from simulated particle trajectories, e.g.\citet{Giacalone1999,Shalchi2009b,Reville2008,Tautz2013,Shalchi2015}.
\citet{Giacalone1999} examine weak and intermediate turbulence with
isotropic and composite geometry (having equal correlation length
in all directions). They find good agreement with QLT for diffusion
parallel to $\boldsymbol{B}_{0}$, with an energy-independent difference
of a factor of $10^{-2}$ from the QLT prediction for perpendicular
diffusion. \citet{Pommois2007} vary the ratio of parallel and perpendicular
magnetic correlation lengths $\frac{l_{\mathrm{cor}\parallel}}{l_{\mathrm{cor}\perp}}$
for fixed values of $\delta B/B_{0}$ weak and intermediate turbulence
and find a variety of distinct transport regimes emerge as a function
of Kubo number\footnote{Here we follow the convention from \citet{Pommois2007}, however this
is greater by a factor of $\sqrt{2}$than the alternative form used
in e.g. \citet{Shalchi2015}.} $K=\frac{\delta B}{B_{0}}\frac{l_{\mathrm{cor}\parallel}}{l_{\mathrm{cor}\perp}}$.
In particular they find anomalous diffusion for slab and isotropic
turbulence, but non-anomalous diffusion for $2d$ wave geometry.
\citet{Shalchi2015} provides a theoretical classification of some
of these regimes in the limit of large or vanishing Kubo number. In
our work the Kubo number is either $0$ (in the case of $r_{\text{\ensuremath{\mathrm{slab}}}}=1$)
or equal to $\delta B/B_{0}$, since both correlation lengths are
equal. In this context our results may explain why varying the Kubo
number is observed by \citet{Pommois2007} to change the anomalous
index associated with the diffusion.

It is also possible to choose other length scales in place of $r_{g}$
where other processes dominate, e.g. turbulence correlation length
or vortex scale \citep{Vladimirov2009a}.

\subsection{Justifications for the Bohm Approximation}

It is commonly accepted that Bohm diffusion is a heuristic, currently
without a rigorous physical derivation \citep{Krall1973,Casse2002},
but it is nevertheless widely used in astrophysics under various justifications.
There is analytic \citep{Shalchi2009b} and numerical \citep{Caprioli2014}
evidence that it is a useful approximation for turbulence levels close
to unity, this is in agreement with the results presented above. In
\citet*{Vladimirov06} it is claimed that a more physically realistic
treatment is necessary, but that a better model of mean free paths
for strong turbulence is analytically intractable (see \citet{Bykovand1992}).
They claim their results are not especially sensitive to the diffusion
model and on that basis it is valid to use the simpler Bohm prescription.
As we have argued however, the transport can vary greatly between
different turbulence levels, and if a simulation is to track particles
and waves across the many orders of magnitude in energy that are required
for DSA then it must account for these differences.

In order to participate in DSA a particle must diffuse efficiently
enough that the shock does not outrun it. For this it is necessary
that the turbulence be strong, or that the diffusion be at least as
strong as Bohm \citep{Achterberg2001}. This has tentative observational
support in SN1006 \citep{Allen2008}. In the context of our results
this would imply that the turbulence level in this environment is
very close to unity.

\subsection{Relevance to Astrophysical Environments}

\begin{table*}
\begin{centering}
\begin{tabular}{cccccc}
\toprule 
 & \multicolumn{2}{c}{SNR} & \multicolumn{2}{c}{GRB\textsuperscript{\ref{fn:piran}}} & AGN \textsuperscript{\ref{fn:dimatteo}}\tabularnewline
\midrule
\midrule 
 & Max & Mean & Internal Shocks & Afterglow\textsuperscript{\ref{fn:assume-jet}} & Flare\tabularnewline
\midrule
$B_{0}$ & $100\mathrm{\mu G}$\textsuperscript{\ref{fn:caprioli}}  & $5\mathrm{\mu G}$ \textsuperscript{\ref{fn:reynold}\ref{fn:woolsey}} & $10^{6}\mathrm{G}$ & $1\mathrm{G}$ & $10^{3}\mathrm{G}$\tabularnewline
\midrule
$\delta B/B$ & $5$\textsuperscript{\ref{fn:bell04}} & $1$\textsuperscript{\ref{fn:bell04}} &  & $5$\textsuperscript{\ref{fn:miz}} & \tabularnewline
\midrule
$E$ & $100\mathrm{TeV}$\textsuperscript{\ref{fn:reynold}} & $10\mathrm{MeV}$\textsuperscript{\ref{fn:guo}} & $100\mathrm{MeV}$ & $100\mathrm{GeV}$ & $100\mathrm{keV}$\tabularnewline
\midrule
$\gamma$ & $10^{5}$ \textsuperscript{\ref{fn:reynold}} & $1.01$\textsuperscript{\ref{fn:guo}} & $1.1$ & $100$ & $\gtrsim1$\tabularnewline
\midrule
$t_{g}$ & $10^{6}\mathrm{s}$ & $10^{2}\mathrm{s}$ & $10^{-10}\mathrm{s}$ & $10^{-2}\mathrm{s}$ & $10^{-6}\mathrm{s}$\tabularnewline
\midrule
$r_{g}$(at $\mu=1$) & $10^{14}\mathrm{m}$  & $10^{5}\mathrm{m}$ & $10^{-2}\mathrm{m}$ & $10^{6}\mathrm{m}$ & $1\mathrm{m}$\tabularnewline
\bottomrule
\end{tabular}
\par\end{centering}
\begin{inparaenum}[\expandafter \textsuperscript a]
\item \citet{DiMatteo1998}\label{fn:dimatteo}
\item \citet{Bell2004}\label{fn:bell04}
\item \citet{Piran2005} (assuming equal proton and electron temperatures)
\label{fn:piran}
\item \citet{Caprioli2009}\label{fn:caprioli}
\item \citet{Reynolds1999}\label{fn:reynold} 
\item \citet{Woolsey2001}\label{fn:woolsey}
\item \citet{Mizuno2011}\label{fn:miz}
\item \citet{Guo2014}\label{fn:guo}
\item (assuming upstream protons entering shock with $\gamma=100$)\label{fn:assume-jet}
\end{inparaenum}
\begin{raggedright}
\caption{\label{tab:Approximate-typical-astrophyisca}Approximate typical astrophysical
parameters for supernova remnants (SNRs), gamma-ray bursts (GRBs)
and active galactic nuclei (AGNs). Rows respectively correspond to:
background magnetic field strength, magnetic turbulence ratio, average
fluid-frame proton energy and Lorentz factor, gyrotime and gyroradius.
Blank cells indicate data was unavailable.}
\par\end{raggedright}
\raggedright{}
\end{table*}

\Tabref{Approximate-typical-astrophyisca} gives estimates of the
relevant parameters for protons in some candidate accelerator environments.
The choice of $\Delta t$ ultimately determines what phenomena can
be treated by modelling a system using \eqref{mufokkerplanck}, or
equivalently the minimum timescale that can be resolved by any simulation
using $D_{{\mu\mu}}$ as a parameter. There are several such timescales
which may be relevant. Two natural choices are: the gyrotime of the
particle $t_{g}$ and the wave crossing time $t_{w}=\frac{2\pi}{k_{max}}$,
where $k_{\mathrm{max}}$ is the largest wavenumber in the spectrum.
However $t_{g}$ is too short to average out nonresonant interactions,
and \citet{Vladimirov2009} show that $k_{\mathrm{max}}$ may be as
large as $10^{2}/r_{g}$ making $t_{w}$ even shorter. 

The physical significance of the $\Delta t$ is manifest in microphysical
processes which may play a role in particle acceleration. The growth
time of relevant plasma instabilities such as Bell \citep{Reville2008,Bai2014}
and Weibel \citep{Schlickeiser03} must be resolved by the diffusive
timestep of a multi-physics simulation. This is because these are
mechanisms by which anisotropies feed the magnetic turbulence and
so are tightly coupled to diffusion.
\begin{center}
\begin{table*}
\begin{centering}
\par\end{centering}
\begin{centering}
\begin{tabular}{llccccc}
\toprule 
\textbf{Timescale} &  &  & SNR & GRB Jet\textsuperscript{\ref{fn:piran-1}} & GRB Afterglow & AGN\textsuperscript{\ref{enu:dimat}}\tabularnewline
\midrule
\midrule 
Gyration &  & $t_{g}=\frac{2\pi\gamma m}{eB_{0}}$ & $10^{2}\mathrm{s}$ & $10^{-10}\mathrm{s}$ & $10^{-2}\mathrm{s}$ & $10^{-6}\mathrm{s}$\tabularnewline
Diffusion &  & $t_{D}=1/D_{\mu\mu}$ & $10^{1}\mathrm{s}$ &  & $10^{0}\mathrm{s}$ & \tabularnewline
Scattering &  & $t_{s}=\int_{0}^{\infty}C(\tau)\mathrm{d}\tau$ & $10^{3}\mathrm{s}$ &  & $10^{1}\mathrm{s}$ & \tabularnewline
Wave Crossing &  & $t_{w}=\frac{1}{\mu v}\frac{2\pi}{k_{\mathrm{max}}}$ &  & $10^{-8}\mathrm{s}$ & $10^{-4}\mathrm{s}$ & \tabularnewline
Alfvén Crossing\textsuperscript{} &  & $t_{A}=L_{\mathrm{system}}/v_{A}$ &  & $10^{3}\mathrm{s}$ & $10^{7}\mathrm{s}$ & $10^{3}\mathrm{s}$\tabularnewline
\bottomrule
\end{tabular}
\par\end{centering}
\begin{inparaenum}[\expandafter \textsuperscript a]
\item \citet{Piran2005} \label{fn:piran-1}
\item \citet{DiMatteo1998}\label{enu:dimat}
\end{inparaenum}
\caption{\label{tab:Table-of-relevant}Table of order-of-magnitude estimates
of relevant time scales. Each of these provides a upper bound for
a simulation timestep $\Delta t$. Diffusion times are estimates using
the data from \figref{time-test}, and scattering times from \figref{Scattering-time-as},
both assuming $r_{\text{slab}}=0.2$. Blank cells indicate data was
unavailable. }
\end{table*}
\par\end{center}

In the presented simulations we have restricted our attention to nonrelativistic
particle velocities, which is justifiable at least for the majority
of particles (see \tabref{Approximate-typical-astrophyisca}) but
not for the high-energy cosmic-ray tail of the energy distribution.
Furthermore since the Lorentz force law (\eqref{lorentz-cov}) is
manifestly covariant, the model considered in this work is applicable
in principle to particles of relativistic energies, with the caveat
that the resonance condition $k_{\text{max}}>1/r_{g}>k_{\text{min}}$
will not satisfied for particles of very large gyroradius. The limiting
wavelength $2\pi/k_{\text{min}}$ may be as large as the system size
$L$ but, depending on the turbulence generation mechanism, there
may be very little magnetic energy available at this scale. Outside
of the test-particle approximation, if a large share of the plasma's
energy is fed into the cosmic ray current, then the self-generated
turbulence may increase in wavelength along with the typical gyroradius
of the particles, and so maintain the resonance condition. Furthermore
as $B$ becomes large the Alfvén speed $v_{A}$ approaches $c$ and
the associated electric field $E$ becomes non-negligible. In this
case particles will gain kinetic energy $T$ via the time component
of \eqref{lorentz-cov}, $\partial_{t}T=ev\cdot E$. While this situation
may well arise in acceleration environments, it significantly complicates
the model and is deferred to future work.

\section{Conclusions}

\label{sec:Conclusions}

In this paper we have shown that, while a comprehensive model of cosmic
ray transport in accelerators is necessary for understanding the origins
of high-energy cosmic rays, existing diffusion models are limited
and may not cover some relevant ranges of parameters. This is because
current analytic approaches (QLT, Bohm) rely on approximations which
are invalid in important turbulence regimes.

The applicability of the diffusion model depends on the timestep/measuring
time $\Delta t$, the choice of which depends on several factors.
It is bounded below by the wave crossing time, the gyrotime, and also
the relevant dynamical timescale for other relevant phenomena (e.g.
plasma instabilities) and is bounded above by the diffusion time,
as demonstrated in \subsecref{choiceofdeltat}. For strong turbulence
therefore, there may be no region in which a valid $\Delta t$ exists.
In the absence of such a $\Delta t$ it is not meaningful to treat
the problem as diffusive and more sophisticated models, e.g. anomalous
diffusion, must be used. To this end we have measured the anomalous
diffusion exponent $\tilde{\alpha}\left(\delta B\right)$ (\figref{alpha-bohm-expo}).

We find that $\tilde{\alpha}\approx0$ at low turbulence levels as
expected from quasilinear theory. Its value then peaks at $\approx3$
for intermediate $\delta B\approx B_{0}$ turbulence and then settles
to $0.5<\tilde{\alpha}<1$. 

The Bohm approximation, while generally applied for its convenience
has been shown to be generally inapplicable to the case of diffusion
in collisionless plasmas of the type described here. As we show, for
environments with intermediate turbulence, the heuristic ``Bohm-type''
models of equation (\ref{eq:bohm-like form}) do not accurately describe
the observed dependence of scattering time on turbulence level. This
is because they necassarily cannot capture the varying $\alpha$ seen
in our simulation results, hence we propose that a more comprehensive
model encapsulating the turbulence dependence is necessary, e.g. a
model of the form:
\[
\lambda_{\mathrm{mfp}}=\eta r_{g}^{\alpha\left(\delta B\right)}
\]

Where the function $\alpha\left(\delta B\right)$ is determined beforehand
in a numerical simulation which includes the details of the scattering
microphysics, as we have done in this work. We have presented \eqref{fitted_model}
as a initial realisation of such a function.

In this work we do not treat feedback from particles to waves. However
this effect can be significant when treating shock dynamics\citep{Caprioli2009,Allen2008},
in particular the acceleration process may dramatically change the
shape of the wave spectrum (see e.g. \citet{Vladimirov2009}). This
will be included in a future work. 

\acknowledgements{}

This research was partially supported by the European Union Seventh
Framework Programme (FP7/2007-2013) under grant agreement ${\rm n}^{\circ}$
618499. The authors also wish to acknowledge the \foreignlanguage{british}{DJEI}/DES/SFI/HEA
Irish \foreignlanguage{british}{Centre} for High-End Computing (ICHEC)
for the provision of computational facilities. We would also like
to thank M. O'Riordan and L. Chen for useful discussions.

\newpage{}

\appendix
\bibliographystyle{apj}
\bibliography{one}

\section{Calculation of Scattering Time}

\label{sec:Calculation-of-Scattering}In order to calculate $t_{s}$
we use a method similar to that of \citet{Casse2002}. While a classical
particle undergoing a deterministic process has its past and future
fully determined by its instantaneous position and velocity, a particle
undergoing stochastic diffusion gradually ``forgets'' its past state.
This loss of information can be quantified, for $\mu\left(t\right)$
real, using the so-called \emph{autocorrelation function},

\begin{equation}
C_{\mu}\left(t,\tau\right)=\frac{\left\langle \mu\left(t+\tau\right)\mu\left(t\right)\right\rangle }{\left\langle \mu\left(t\right)^{2}\right\rangle }\label{eq:cttau}
\end{equation}
where chevrons indicate any of three types of average: magnetic turbulence
ensemble, chaotic motion ensemble, or temporal. If either of the former
two types of chaos are ergodic, then they are equivalent to the temporal
average, in which case \eqref{cttau} simplifies to 
\begin{align*}
C_{\mu}\left(\tau\right) & =\frac{\left\langle \mu\left(t+\tau\right)\mu\left(t\right)\right\rangle _{t}}{\left\langle \mu\left(t\right)^{2}\right\rangle _{t}}\\
 & =\frac{\int_{-\infty}^{\infty}\mu\left(t+\tau\right)\mu\left(t\right)\,\mathrm{d}t}{\int_{-\infty}^{\infty}\mu\left(t\right)^{2}\,\mathrm{d}t}\\
 & =\frac{\mu\left(\tau\right)*\mu\left(-\tau\right)}{\left\Vert \mu\right\Vert ^{2}}
\end{align*}
where $*$ indicates convolution and $\left\Vert \cdot\right\Vert $
is the standard norm. If we now apply the convolution theorem and
omit the $\tau$ for clarity,
\begin{align*}
\mathcal{F}C_{\mu} & =\left(\mathcal{F}\mu\right)\left(\mathcal{F}\mu\right)^{\ast}/\left\Vert \mu\right\Vert ^{2}\\
C_{\mu} & =\mathcal{F}^{-1}\left|\mathcal{F}\mu\right|^{2}/\left\Vert \mu\right\Vert ^{2}
\end{align*}
where $\mathcal{F}$ is the Fourier transform, asterisk denotes complex
conjugate, and $\left|\cdot\right|$ gives the magnitude of a complex
number. This is known as Wiener-Khinchin form of the autocorrelation,
and is the form used in this work for the purposes of numerical calculation
since it is much more computationally efficient than the convolution
form.

Formally the scattering time is then given by 
\[
t_{s}=\int_{0}^{\infty}C_{\mu}\left(\tau\right)\mathrm{d}\tau.
\]
In the case of a classical Gaussian diffusion we expect the autocorrelation
to decay exponentially, $C_{\mu}\left(\tau\right)=e^{-\alpha\tau}$
for some real parameter $\alpha$ and so we simply find $t_{s}=1/\alpha$.
In the present work however, $C_{\mu}\left(\tau\right)$ is found
to be highly oscillatory. Numerical integrals of this type are notoriously
difficult to perform reliably, and for this reason we make the following
simplifying assumption, that the numerical $C_{\mu}$s measured in
this work are the product of various oscillating signals, and an exponentially
decaying exponential envelope, so that $C_{\mu}\left(\tau\right)=\Re e^{t\left(\alpha+\beta i\right)}$,
where $\beta$ is an ignorable real parameter, and $\Re$ takes the
real part of a complex number. Here the decay constant $\alpha$ is
the reciprocal of $t_{s}$ as above and this is the value that is
presented in our results.

Care must be taken when using this method, as the finite length of
the $C_{\mu}$ measurement means that even in the absence of diffusion
$C_{\mu}$ will exhibit a $1/\tau$ envelope behaviour, with a slope
of $-1/t_{\max}$, corresponding to a best fit scattering time of
$t_{\mathrm{max}}\left(1-e^{-1}\right)$. To avoid measuring this
spurious signal and have the real diffusion dominate, the $t_{\mathrm{max}}$
must be set to at least $2t_{s}$. Since $t_{s}$ cannot be known
beforehand, this is achieved iteratively by increasing $t_{\mathrm{max}}$
until a good fit is obtained.

\end{document}